\documentclass[aps,prb,twocolumn,groupedaddress]{revtex4-1}

\usepackage[utf8]{inputenc}
\usepackage{graphicx}
\usepackage{bm}
\usepackage{amsmath}
\usepackage{amssymb}
\usepackage{siunitx}
\usepackage{mathtools}
\usepackage{hyperref}
\usepackage{physics}
\hypersetup{colorlinks=true,urlcolor=black,linkcolor=black,citecolor=black}

\newcommand\plot[3]{\begin{figure} \includegraphics[#2]{plot_#1}\caption{\label{fig:#1} #3}\end{figure}}
\newcommand\fig[3]{\begin{figure} \includegraphics[#2]{fig_#1}\caption{\label{fig:#1} #3}\end{figure}}
\newcommand\mat[1]{\vb{#1}}
\newcommand\vek[1]{\vb{#1}}
\newcommand\matrdd[4]{\begin{pmatrix} #1 & #2 \\ #3 & #4 \end{pmatrix}}

\newcommand\defeq{\vcentcolon =}

\newcommand\ladd{^{\phantom{\dag}}}
\newcommand\ecom{\,,}
\newcommand\edot{\,.}
\newcommand\ead[1]{\mathrm{e}^{#1}}
\newcommand\D{\mathop{}\!\mathrm{d}}
\newcommand{\gvek}[1]{\vb*{#1}}
\newcommand{\rvek}[1]{\vek{\mathcal{#1}}}
\newcommand\pdim{\mathcal{P}}
\newcommand\xdim{\mathcal{X}}
\newcommand\qdim{\mathcal{Q}}
\newcommand\edim{\mathcal{E}}
\newcommand\mdim{\mathcal{M}}

\newcommand\ddim{\mathcal{D}}
\newcommand\la{\vb{n}}
\newcommand\lb{\vb{n}'}
\renewcommand\ln{\vb{n}}
\newcommand\lm{\vb{m}}

\begin{document}

\title{Thermoelectric transport coefficients of a Dirac electron gas in high magnetic fields}



\author{Viktor Könye}
\author{Masao Ogata}
\affiliation{Department of Physics, The University of Tokyo, Bunkyo-ku, Tokyo 113-0033, Japan}

\date{\today}

\begin{abstract}
We study the thermoelectric transport properties of a three-dimensional massive relativistic fermion gas with screened Coulomb impurities in high magnetic fields where only the lowest Landau levels contribute to the transport. Our results can be applied to experimental results of gapless and gapped Dirac materials. We focus on the effects of the mass term and we show the main differences that arise compared to the massless Dirac fermions. The different behavior is shown to be relevant at higher magnetic fields. The calculations are performed in the framework of the linear response theory using the exact quantum mechanical solution of the system in a constant magnetic field. We prove that the Mott formula and the Wiedemann-Franz law are valid at low temperatures and use them to calculate the thermoelectric transport coefficients. We show that the temperature range where the low temperature approximation is valid increases with increasing magnetic fields. The magnetic field dependence of measurable quantities (i.e. conductivity, Seebeck coefficient, Nernst coefficient and thermal conductivity) strongly depend on the magnetic field dependence of the scattering rate, thus the result relies on the proper treatment of the impurities. In this work they are included through the first Born approximation using screened charged impurities as impurity potential. We show that the electric conductivity does not change qualitatively in the case of finite mass term. On the other hand we find that the mass term causes significantly different behavior in the Seebeck and Nernst coefficients.
\end{abstract}


\maketitle

\section{Introduction}

  The Dirac equation\cite{Dirac1928} developed almost a century ago, played a very important role in understanding relativistic fermions in particle physics. In recent years it is extensively used in the field of condensed matter physics to describe three-dimensional Dirac materials which were found to have relativistic fermions as low energy excitations (for recent reviews see Refs. \onlinecite{Armitage2018,Bernevig2018,Fuseya2015}).
 
Materials exhibiting massless fermions are topologically classified as Dirac or Weyl semimetals\cite{Chiu2016}. Experimentally studied gapless Dirac materials include $\mathrm{Cd}_3\mathrm{As}_2$\cite{Crassee2018,Liu2014,Borisenko2014}, $\mathrm{Na}_3\mathrm{Bi}$\cite{Liu2014a} and $\mathrm{TaAs}$\cite{Xu2015}. Materials with massive fermions are called gapped Dirac materials of which several three dimensional candidates were found experimentally\cite{Hsieh2008,Orlita2015,Chen2017,Yuan2017,Suetsugu2018}. Furthermore, two-dimensional Dirac fermions can be found in several quasi-two-dimensional materials. The classical example for massless fermions is graphene\cite{Peres2009}. For the massive case an extensively studied material is the bilayer graphene\cite{McCann2013,McCann2006,Nam2010}.

Three-dimensional Dirac materials show several interesting transport properties (for recent reviews see Refs. \onlinecite{Gorbar2018,Wang2017}). Under an external magnetic field some of these are the chiral anomaly\cite{Nielsen1983}  and as a consequence negative magnetoresistance\cite{Huang2015} and a non-saturating linear magnetoresistance\cite{He2014,Liang2015,Narayanan2015,Suetsugu2018,Leahy2018}. Furthermore, several thermoelectric experiments were carried out in an external magnetic field\cite{Xiang2017,Liang2013,Caglieris2018,Liang2017,Gooth2018}. The thermopower was found to be increasing linearly with the magnetic field\cite{Liang2013}. At lower fields the Nernst coefficient shows an anomalous behavior\cite{Liang2017,Caglieris2018}.

The most simple continuum Hamiltonian to describe Dirac materials is the $4\times4$ Dirac Hamiltonian\cite{Wolff1964,Zhang2009,Fuseya2015} with effective values for the speed of light and mass of electron. This model is able to reproduce many properties that were experimentally observed. In the case of zero effective mass we get the Weyl Hamiltonian suitable to describe Dirac and Weyl semimetals\cite{Young2012,Wang2013a,Kariyado2012,Kariyado2017}.

Using this continuum model previous theoretical papers investigate the magnetoconductivity in the massless\cite{Abrikosov1998,Xiao2017,Klier2017a} and massive\cite{Konye2018,Proskurin2015,Wang2018} cases. The experimentally seen linear magnetoresistance is recovered\cite{Abrikosov1998,Xiao2017,Klier2017a,Konye2018} only if screened Coulomb impurities are used. These studies use the linear response theory using the Born approximation\cite{Shon1998a} to evaluate the self-energy. The vertex correction in the massless case was investigated in Ref. \onlinecite{Klier2015a}, where they found that close to the Weyl point the effect of the vertex correction is negligible.

Several papers investigate the thermoelectric coefficients of Dirac materials in the zero magnetic field case\cite{Peng2016,Saha2018,McCormick2017,Gorbar2017b}. The finite Berry curvature of Weyl semimetals causes the anomalous Nernst and thermal Hall effects even in no magnetic field\cite{McCormick2017}. In the case of low magnetic fields studies using the semiclassical Boltzmann approach can be found in Refs. \onlinecite{Saha2018,Lundgren2014,Sharma2017}. For the case of high magnetic fields, the Seebeck coefficient for the Weyl Hamiltonian was studied in Ref. \onlinecite{Skinner2018} and was found to be linear and non-saturating at high fields. Their calculation is based on expressing the thermopower using the entropy density. In this paper, we study the thermoelectric transport in high fields using the linear response theory.

For the transport coefficients we follow Luttinger's argument\cite{Luttinger1964}. As it was discussed previously the correct treatment of the transport coefficients need the magnetization and the so called energy magnetization\cite{Smrcka1977,Qin2011}. Otherwise divergences can appear in the off-diagonal components of the transport coefficient tensors.

In usual systems the thermoelectric transport coefficients can be expressed using only the zero temperature conductivity as a function of the chemical potential. These relations, which we called Sommerfeld-Bethe (SB) relations\cite{OgataFukuyama,Matsuura2019}, can be obtained from the Boltzmann transport equation. It was Jonson and Mahan\cite{Jonson1980} who firstly showed these relations microscopically for the case of a single-band Hamiltonian with static random potentials and static phonons. They also discussed the violation of the SB relations in the presence of electron-phonon interaction. The validity of the SB relations has been discussed in the presence of mutual interactions between electrons\cite{OgataFukuyama,Kontani2003}.

In the present paper, we first describe the system in an external magnetic field. We calculate the chemical potential, the screening and the scattering rate and we give their asymptotic behavior at high magnetic fields based on our results in Ref. \onlinecite{Konye2018}. Then, we move on to the calculation of the thermoelectric transport coefficients. We study the magnetic field and mass term dependence of the transverse components of the conductivity, thermal conductivity, Seebeck and Nernst coefficients. In the Appendices we discuss the details of the formalism used to calculate the transport coefficients. We prove that the SB relations can be used for our system, thus the Mott formula and Wiedemann-Franz law are valid. We clarify the temperature range where the low temperature approximation can be used and show that this range increases with the magnetic field.

\section{Model}

  The studied system is a three-dimensional relativistic electron gas in a constant magnetic field. The system is the same as in Ref. \onlinecite{Konye2018}. In this section we list only what is important to understand the later sections (for more details see Ref. \onlinecite{Konye2018}). The single-particle Hamiltonian is:
  \begin{equation}
    \label{eq:ham}
    \mat{H}_D\defeq\mat{\gamma}^0\left[\sum\limits_{i=1}^3v\mat{\gamma}^i\left(p_i+eA_i\right)+\Delta\right]\ecom
  \end{equation}
  Where $\mat{\gamma}^\mu$ are the Dirac matrices and $\Delta$ is the mass term. The external uniform magnetic field points in the $z$ direction ($\vek{A}=(0,Bx,0)$). From now on we use units where $v=1$, $\hbar=1$ and $k_B=1$. The Landau levels are\cite{Konye2018}:
  \begin{equation}
    \label{eq:landau}
    E_{n\lambda s}(p_z)=\lambda\sqrt{2neB+\Delta^2+p_z^2}\ecom
  \end{equation}
where $n=0,1,2,\dots$ is the Landau index, $\lambda=\pm1$ represents the band index and $s=\pm1$ represents the two-fold degeneracy (for $n\neq0$ levels). The eigenstates are\cite{Kaminker1981,Konye2018}:
  \begin{equation}
    \label{eq:eigs1}
    \ket{\vek{\Phi}_{n\lambda s}} = \begin{pmatrix*}[l] \phantom{-\lambda} u_{n,\lambda,s}&\ket{n-1} \\ \phantom{-\lambda}\llap{$-s$}u_{n,\lambda,-s}&\ket{n} \\ \phantom{-\lambda}\llap{$s\lambda$} u_{n,-\lambda,s}&\ket{n-1} \\ -\lambda u_{n,-\lambda,-s}&\ket{n}  \end{pmatrix*}\ecom
  \end{equation}
where in the case of $n=0$, $s=-\mathrm{sgn}(p_z)$ and $u_{n\lambda s}$ is given by:
  \begin{equation}
    \label{eq:un}
    u_{n\lambda s}=\frac{1}{2}\sqrt{\left( 1+\frac{sp_z}{\sqrt{E_n^2-\Delta^2}}   \right)\left( 1+\lambda\frac{\Delta}{E_n}    \right)}\ecom
  \end{equation}
with $E_n\equiv E_{n11}(p_z)$. The quantum numbers of the system are $\la\equiv(n,\lambda,s,p_z,p_y)$. The Landau levels only depend on $n$, $\lambda$ and $p_z$. It is useful to introduce the magnetic length as:
\begin{equation}
  \ell_B\defeq\sqrt{\frac{\hbar}{eB}}\edot
\end{equation}
 Each Landau level is $L^2/2\pi\ell_B^2$-fold degenerate in $p_y$ ($L$ is the length of the system) and twofold degenerate in $s$ (for $n\neq0$). The wave function of state $\ket{n}$ can be expressed with the orthonormal Hermite-functions:
  \begin{equation}
    \braket{x}{n} = \frac{i^n}{L}h_n(x+\ell_B^2p_y;\ell_B)\ead{ip_y y}\ead{ip_z z}\ecom
  \end{equation}
  \begin{equation}
  h_n(x;\ell_B) \defeq \frac{(\ell_B^2\pi)^{-1/4}}{\sqrt{2^nn!}}\exp(-\frac{x^2}{2\ell_B^2})H_n\left(\frac{x}{\ell_B}\right)\ecom
  \end{equation}
 where $H_n(x)$ are the Hermite-polynomials.

The Green's function for the $\vek{H}_D$ system is:
\begin{equation}
\mat{G}_D(i\omega_m) = \sum\limits_{\la}\frac{\ket{\gvek{\Phi}_{\la}}\bra{\gvek{\Phi}_{\la}}}{i\omega_m+\mu-E_{\la}}\edot
\end{equation}
In later sections we will use three different representations of the Green's function. First, the coordinate representation:
\begin{equation}
\mat{G}_D(\vek{x},\vek{x}',i\omega_m) = \sum\limits_{\la}\frac{\gvek{\phi}_{\la}\ladd(\vek{x})\gvek{\phi}_{\la}^\dag(\vek{x}')}{i\omega_m+\mu-E_{\la}}\ecom
\end{equation}
where $\vek{\phi}_{\la}(\vek{x})\defeq\braket{\vek{x}}{\vek{\Phi}_{\la}}$. Second, the momentum representation:
\begin{equation}
\mat{G}^{D}_{\vek{k}\vek{k}'}(i\omega_m) = \sum\limits_{\la}\frac{\gvek{\phi}_{\la\vek{k}}\ladd\gvek{\phi}_{\la\vek{k}'}^\dag}{i\omega_m+\mu-E_{\la}}\ecom
\end{equation}
where $\gvek{\phi}_{\la\vek{k}}\ladd=\int\D^3x\mathrm{e}^{-i\vek{kx}}\gvek{\phi}_{\la}\ladd(\vek{x})$. Finally, the Landau level representation:
\begin{equation}
G^{D}_{\lb\la}(i\omega_m) = \frac{\delta_{\la\lb}}{i\omega_m+\mu-E_{\la}}\edot
\end{equation}
The connections between these representations are:
\begin{align}
\mat{G}_{\vek{k}\vek{k}'}(i\omega_m) &= \int\D^3 x \D^3 x' \ead{-i\vek{k}\vek{x}}\mat{G}(\vek{x},\vek{x}',i\omega_m)\ead{i\vek{k}'\vek{x}'}\ecom\\
\mat{G}(\vek{x},\vek{x}',i\omega_m) &= \frac{1}{V^2}\sum\limits_{\vek{k},\vek{k}'}\ead{i\vek{kx}}\mat{G}_{\vek{kk}'}(i\omega_m)\ead{-i\vek{k}'\vek{x}'}\ecom\\
\mat{G}_{\vek{k}\vek{k}'}(i\omega_m)&= \sum\limits_{\la,\lb} \gvek{\phi}_{\la\vek{k}}G_{\la\lb}(i\omega_m) \gvek{\phi}_{\lb\vek{k}'}^\dag\ecom\\
G_{\la\lb}(i\omega_m) &= \frac{1}{V^2}\sum\limits_{\vek{k},\vek{k}'}\gvek{\phi}_{\la\vek{k}}^\dag \mat{G}_{\vek{k}\vek{k}'}(i\omega_m)\gvek{\phi}_{\lb\vek{k}'}\edot
\end{align}
\section{Impurity Green's function}
\label{sec:impurity}
In this section, we summarize the results in Ref \onlinecite{Konye2018} that are used in the present paper. In particular, we focus on the quantum limit. 

First, the chemical potential as a function of the magnetic field is calculated by fixing the charge carrier density. The chemical potential in the $\ell_B\to0$, $T\to0$ limit is\cite{Konye2018}:
\begin{align}
    \label{eq:chemlim}
    \mu &\sim \sqrt{\Delta_B^2+\Delta^2} & \Delta_B&\defeq2\pi^2n_e\ell_B^2\propto\frac{1}{B}  \edot
\end{align}

The magnetic fields where the chemical potential crosses the Landau levels can be calculated as:
  \begin{align}
    \label{eq:Bpeak}
    eB_m=\left(\frac{\sqrt{2}\pi^2n_e}{A(m)}\right)^{\frac{2}{3}},~A(m)\defeq\sum\limits_{n=0}^m(2-\delta_{n0})\sqrt{m-n}\ecom
  \end{align}
With this the criteria for the quantum limit can be expressed as:
\begin{equation}
\ell_B^2(\sqrt{2}\pi^2n_e)^{2/3}\ll1\edot
\end{equation}

\sisetup{per-mode=fraction}
Using experimentally realistic parameters such as $n_e\approx10^{18}\si{\per\cubic\centi\metre}$\cite{Liang2015,Suetsugu2018}, $v=\SI{1e6}{\metre\per\second}$\cite{Crassee2018,Liu2014,Borisenko2014,Liang2015}, $\Delta=10-50\si{\milli\electronvolt}$\cite{Fuseya2015,Chen2017} the quantum limit is reached above $B\approx\SI{40}{\tesla}$ and the mass term becomes relevant above $B=850-170\si{\tesla}$.
\sisetup{per-mode=reciprocal}
The chemical potential as a function of the magnetic field is shown in Fig. \ref{fig:chem}
    \plot{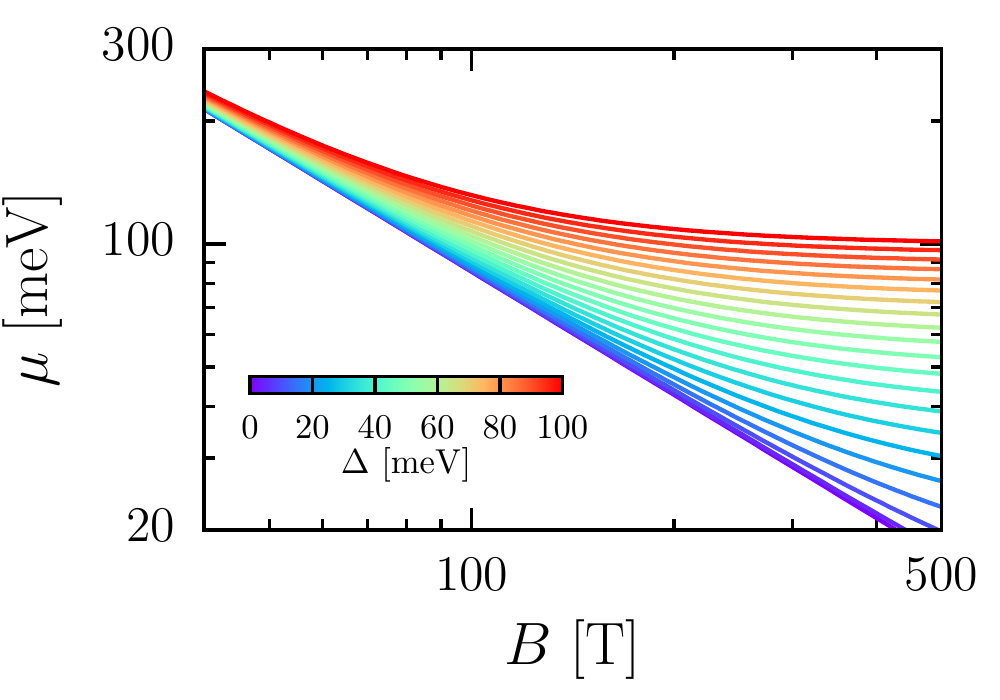}{width=.45\textwidth}{The chemical potential in the quantum limit (\ref{eq:chemlim}) as a function of the magnetic field using different mass terms. The parameters used are: $n_e=10^{18}\si{\per\cubic\centi\metre}$ and $v=10^{6}\si{\metre\per\second}$.}

The impurities are included to the Dirac Hamiltonian in Eq. (\ref{eq:ham}) as:
\begin{equation}
\label{eq:hamwithim}
\vek{H}=\vek{H}_D+\sum\limits_i u(\vek{x}-\vek{x}_i)\ecom
\end{equation}
where $u(\vek{x}-\vek{x}_i)$ is the impurity potential at position $\vek{x}_i$ which is assumed to be uniformly distributed. The self-energy is calculated using the first-order Born approximation as in Refs. \onlinecite{Konye2018,Abrikosov1998}. The scattering rate in the Landau level representation ($\Gamma_{\la}=-\Im{\Sigma_{\la}}$) becomes diagonal and can be expressed as\cite{Konye2018}:
    \begin{widetext}
      \begin{subequations}
      \begin{align}
	\label{eq:gam}
        \Gamma_{n\lambda s}(\edim,\pdim_z,B) &=\frac{n_i\pi}{\ell_B^2}\sum\limits_{\ell=0}^{\left\lfloor\frac{(\edim+\mdim)^2-\mathcal{D}^2}{2} \right\rfloor}\sum\limits_{\substack{\alpha = \pm 1 \\ t=\pm1}} \int\frac{\D \qdim_x\D \qdim_y}{(2\pi)^3}u_{\bm{\qdim}_{\ell\alpha}}^2(B) \left|\frac{\edim+\mdim}{\sqrt{(\edim+\mdim)^2-2\ell-\mathcal{D}^2}} \right|
       \left|F_{n\lambda s}^{\ell\gamma_0t}(\bm{\qdim}_{\ell\alpha},\pdim_z)\right|^2\ecom
      \end{align}
      \begin{align}
	\label{eq:four}
        F_{n\lambda s}^{\ell\gamma t}(\bm{\qdim},\pdim_z) &\defeq \int\D \xdim \gvek{\phi}_{n\lambda s}^\dag(\xdim;0,\pdim_z) \gvek{\phi}_{\ell\gamma t}(\xdim;\qdim_y,\pdim_z-\qdim_z)\ead{i\qdim_x\xdim}\ecom
      \end{align}
      \begin{align}
        \bm{\qdim}_{\ell\pm} &\defeq ({\qdim_x},{\qdim_y},{\qdim_{\ell\pm}})\ecom
        &\qdim_{\ell\pm} &\defeq \pdim_z\pm\sqrt{(\edim+\mdim)^2-2\ell-\ddim^2}\ecom
      \end{align}
      \end{subequations}
    \end{widetext}
where $\gamma_0 = \mathrm{sgn}(\edim+\mdim)$, $\gvek{\phi}_{n\lambda s}(x;p_y,p_z)\defeq\braket{\vek{x}}{\vek{\Phi}_{\vek{n}}}$  and using $\ell_B =\sqrt{\hbar/eB}$:
    \begin{align}
	\label{eq:dimensionless}
       (\bm{\pdim},\bm{\qdim},\mathcal{E},\mathcal{M},\mathcal{D},x) &\defeq\ell_B (\vek{p},\vek{q},\varepsilon,\mu,\Delta,\mathcal{X})\edot
    \end{align}
The summation over $t$ is only for $l\neq0$. In this formula the magnetic field dependence of the scattering rate comes from the explicit $\ell_B$ factor, the impurity potential, $\mdim$ and $\ddim$.

For the impurity potential $u_{\vek{q}}$ we will take into account the screening through the electron-electron interaction. For this the so called Random Phase Approximation\cite{Bruus2004,Mahan2000} (RPA) was used. In the long wavelength limit ($\vek{q},\vek{q}'\to0$) the screened impurity potential becomes\cite{Konye2018}:
  \begin{equation}
    u_{\vek{q}}=\frac{u_i}{q^2+\kappa^2}\ecom
  \end{equation}
where the screening wavelength in the quantum limit becomes\cite{Konye2018}:
  \begin{equation}
  \label{eq:kappalimit}
	  \kappa^2\sim\frac{u_e}{2\pi^2\ell_B^2}\frac{\sqrt{\Delta_B^2+\Delta^2}}{\Delta_B}\propto\begin{cases}B&\Delta_B\gg\Delta\\B^2&\Delta_B\ll\Delta\end{cases}\edot
  \end{equation}
The screening wavelength as a function of the magnetic field is shown in Fig. \ref{fig:screen}
    \plot{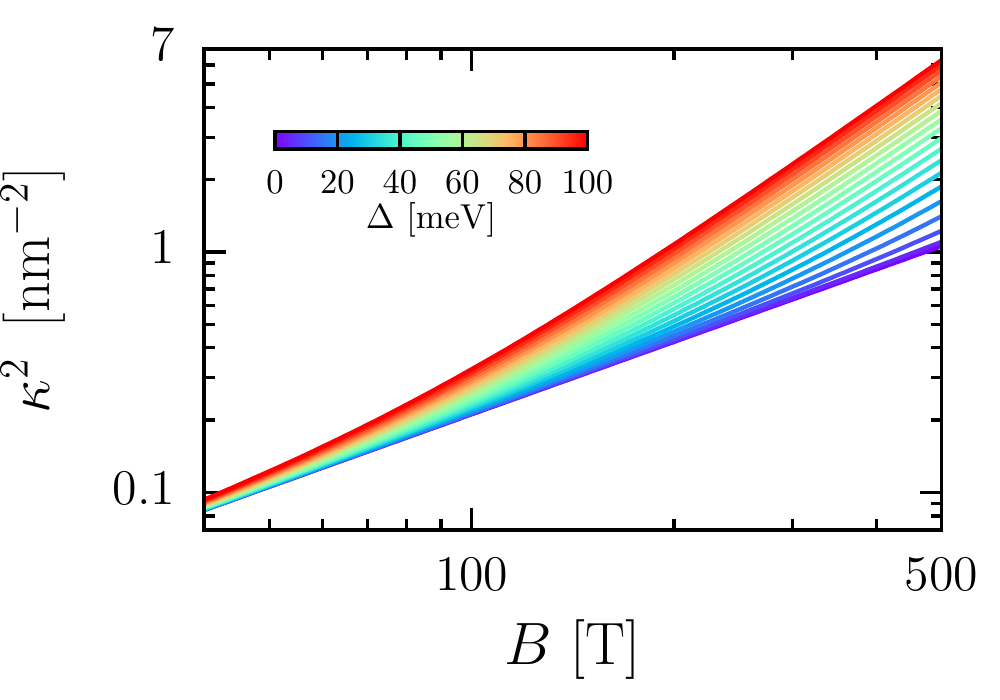}{width=.45\textwidth}{Screening wavenumber in the quantum limit (\ref{eq:kappalimit}) as a function of the magnetic field using different mass terms. The parameters used are: $n_e=10^{18}\si{\per\cubic\centi\metre}$ and $v=10^6\si{\metre\per\second}$.}

The above formulas were derived in Ref. \onlinecite{Konye2018}. Using them we can get a formula for the scattering rate in the quantum limit. In later sections we will see that at low temperatures the important part of the scattering rate is at $\edim=0$. Using the (\ref{eq:kappalimit}) screening in Eq. (\ref{eq:gam}) in the quantum limit $\Gamma_{n\lambda s}\equiv\Gamma_{n\lambda s}(0,\pdim_z,B)$ becomes:

      \begin{align}
	\label{eq:gamlimit}
        \Gamma_{n\lambda s} &=n_i u_i^2 \ell_B^2\hspace{-4pt}\sum\limits_{\alpha = \pm} \hspace{-3pt}\int\hspace{-3pt}\frac{\D \qdim_x\D \qdim_y}{2(2\pi)^2}\frac{ \frac{\sqrt{\ddim_B^2+\ddim^2}}{\ddim_B} \hspace{-3pt}\left|F_{n\lambda s}^{0\gamma_0\tilde{t}}(\bm{\qdim}_{0\alpha},\pdim_z)\right|^2}{\left(\bm{\qdim}_{0\alpha}^2+\frac{u_e}{2\pi^2}\frac{\sqrt{\ddim_B^2+\ddim^2}}{\ddim_B}\right)^2}\edot
      \end{align}

In the $\ell_B\to0$ limit this can be expressed as:
\begin{align}
\Gamma_{n\lambda s}&=n_iu_i^2\ell_B^2\begin{cases} I_{n\lambda s}&\Delta_B\gg\Delta\\\frac{\Delta_B}{\Delta}J_{n\lambda s}&\Delta_B\ll\Delta\\\end{cases}\ecom
\end{align}
\begin{align}
I_{n\lambda s}&\defeq\sum\limits_{\alpha = \pm 1} \int\frac{\D \qdim_x\D \qdim_y}{2(2\pi)^2}\frac{\left|F_{n\lambda s}^{0\gamma_o\tilde{t}}(\bm{\qdim}_{0\alpha},0)\right|^2}{\left(\bm{\qdim}_{0\alpha}^2+\frac{u_e}{2\pi^2}\right)^2}\\
J_{n\lambda s}&\defeq\sum\limits_{\alpha = \pm 1} \int{\D \qdim_x\D \qdim_y}\frac{\pi^2}{2u_e^2}\left|F_{n\lambda s}^{0\gamma_o\tilde{t}}(\bm{\qdim}_{0\alpha},0)\right|^2
\end{align}
The magnetic field dependence is:
\begin{equation}
\label{eq:gambigB}
\Gamma_{n\lambda s}\propto\begin{cases} B^{-1}&\Delta_B\gg\Delta\\B^{-2}&\Delta_B\ll\Delta\\\end{cases}\edot
\end{equation}
The scattering rate calculated numerically from Eq. (\ref{eq:gamlimit}) can be seen in Fig. \ref{fig:Gamma_B}. As we can see the scattering rate depends strongly on the Landau index. At lower fields the mass term only affects the quantitative value of the scattering rate, but the magnetic field dependence is unaffected. At high fields we can see the dependencies described in Eq. (\ref{eq:gambigB}). For higher Landau levels a higher magnetic field is needed to get the asymptotic behavior.
    \plot{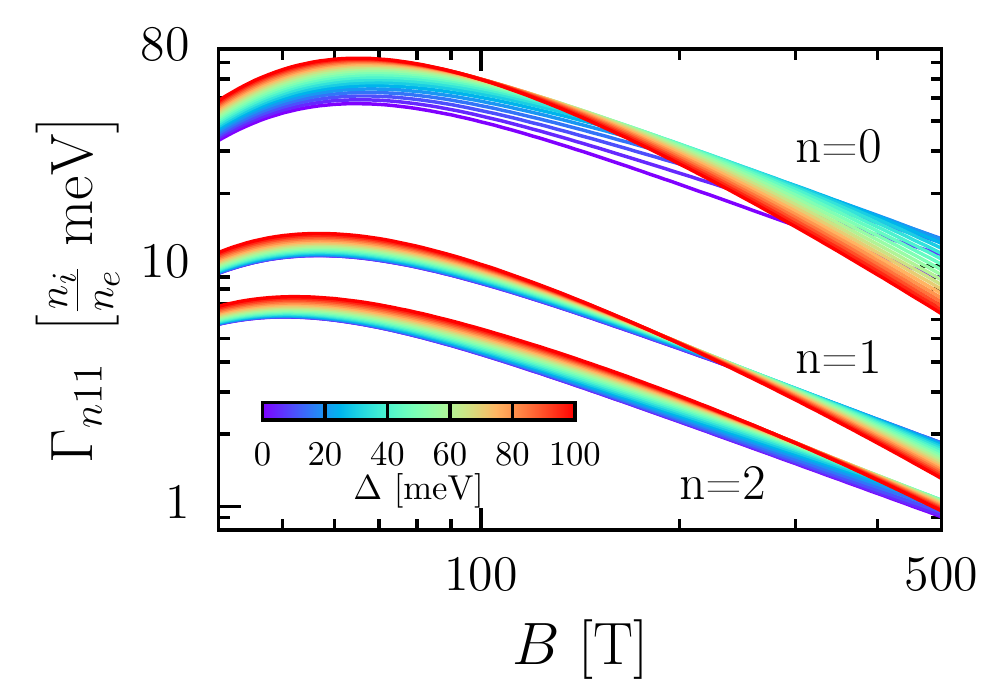}{width=.45\textwidth}{Scattering rate of the $n=0,1,2$ Landau levels in the quantum limit (\ref{eq:gamlimit}) as a function of the magnetic field using different mass terms. The scattering rate is calculated at $\pdim_z=0$ and $\edim=0$. The parameters used are: $n_e=10^{18}\si{\per\cubic\centi\metre}$ and $v=10^6\si{\metre\per\second}$.}

Using the (\ref{eq:gam}) scattering rate the impurity Green's function can be calculated from the Dyson equation as:
  \begin{equation}
    \label{eq:green}
   G_{\la}(i\omega_m) = \frac{1}{i\omega_m+\mu-E_{\la}+i\Gamma_{\la}(i\omega_m)}\edot
  \end{equation}

\section{Transport coefficients}
\label{sec:transport}
\subsection{Calculation of transport coefficients in a magnetic field}
\label{sec:trans}
The many-body Hamilton operator in the presence of external potentials can be expressed as\cite{Luttinger1964}:
\begin{equation}
\mathcal{H}_{\mathrm{tot}}=\int \D^3x h(\vek{x})[1+\psi(\vek{x})]+\varrho_e(\vek{x})\phi(\vek{x})\ecom
\end{equation}
where $\psi$ is the fictitious gravitational potential introduced as the dynamical counterpart of the temperature gradient, $\phi$ is the electric potential, $h$ is the energy density and $\varrho_e$ is the charge density. Using the explanation in Appendix \ref{app:curr} and \ref{app:currex} the single-particle current and energy current operator for the Eq. (\ref{eq:hamwithim}) Hamiltonian can be expressed as:
  \begin{subequations}
\begin{align}
\va{\vek{J}}_1^{\mathrm{tot}}=&\va{\vek{J}}_1+\va{\vek{J}}_1\psi\ecom\\
\va{\mat{J}}_2^{\mathrm{tot}}=&\va{\vek{J}}_{2}-e\va{\vek{J}}_1\phi+\frac{1}{2}\left[\va{\vek{J}}_{2}\psi+\psi\va{\vek{J}}_{2}+\va{\vek{J}}_1\psi\vek{H}_0+\vek{H}_0\psi\va{\vek{J}}_1\right]\ecom
\end{align}
  \end{subequations}
where
\begin{align}
\va{\vek{J}}_1&=\grad_{\vek{p}}\vek{H}=\matrdd{0}{\mat{\sigma}^\alpha}{\mat{\sigma}^\alpha}{0}\ecom&
\va{\vek{J}}_2&=\frac{1}{2}[\va{\vek{J}}\vek{H}+\vek{H}\va{\vek{J}}]\edot
\end{align}
The matrix elements of the current operator ($\va{J}\equiv\va{J}_1$) in the Landau level representation can be calculated as:
  \begin{equation}
    J^{(\alpha)}_{\la\lb} = \int\D^3 x \gvek{\phi}_{\la}^\dag(\vek{x}) \mat{J}_\alpha \gvek{\phi}_{\lb}(\vek{x})\edot
  \end{equation}
Using the eigenstates in Eq. (\ref{eq:eigs1}) the matrix elements are:
  \begin{subequations}
        \label{eq:jmat}
  \begin{align}
        \label{eq:jx}
        J_{\la\lb}^{(x)} &= \phantom{i}\delta_{p_yp_y'}\delta_{p_zp_z'}\delta_{n,n'-1}U_{n\lambda s}^{n'\lambda's'}+(\la\leftrightarrow \lb)\ecom\\
        J_{\la\lb}^{(y)} &= i\delta_{p_yp_y'}\delta_{p_zp_z'}\delta_{n,n'-1}U_{n\lambda s}^{n'\lambda's'}-(\la\leftrightarrow \lb)\ecom\\
	U_{n\lambda s}^{n'\lambda's'} &\defeq -\lambda u_{n,-\lambda,-s}u_{n',\lambda',s'}-ss'\lambda'u_{n,\lambda,-s}u_{n',-\lambda',s'}\edot
   \end{align}
   \end{subequations}

We define the transport coefficients ($\vek{L}_{ij}$) as\cite{Luttinger1964,Smrcka1977,Mahan2000}:
\begin{subequations}
  \begin{align} 
    \vek{j}_1 &= -e^2\vek{L}_{11}\grad\phi+e\vek{L}_{12}\grad\psi\ecom\\
    \vek{j}_2 &= \phantom{-}e\phantom{^2}\vek{L}_{21}\grad\phi  - \phantom{e}\vek{L}_{22}\grad\psi\ecom
  \end{align}
\end{subequations}
where we separated the elementary charge ($e$) from the usual definitions. In Appendix \ref{app:trans} we show that the transport coefficients in the framework of linear response theory can be expressed as:
\begin{equation}
\label{eq:appexplained}
\mat{L}_{ij}(T,\mu)=-\int\dd \varepsilon \dv{f(\varepsilon-\mu)}{\varepsilon}\mat{L}_{ij}(0,\varepsilon)\edot
\end{equation}

The zero temperature conductivity $e^2\vek{L}_{11}\equiv\vb*{\sigma}$ can be expressed as:
\begin{align}
\nonumber
L^{11}_{\alpha\beta}(0,\varepsilon)&=\frac{1}{\pi V}\sum\limits_{a,b} \bigg[ \Re{J^{(\alpha)}_{ab}J^{(\beta)}_{ba}}\Im G^R_b(\varepsilon) \Im G^R_a(\varepsilon) +\\
\label{eq:L11full}
 &+ \Im{J^{(\alpha)}_{ab}J^{(\beta)}_{ba}}\int\limits_{-\infty}^\varepsilon \D\xi 2\partial_\xi \Re G_b^R(\xi)\Im G^R_a(\xi)\bigg]\edot
\end{align}
In the appendix the derivation of this formula assumes that the eigenvalue problem of the full Hamiltonian (\ref{eq:hamwithim}) is known. In our case only the $\vek{H}_D$ part can be solved, and we treat the impurities as perturbation. In Appendix \ref{app:vertex} we explain how this can be done including vertex corrections. 

In Appendix \ref{app:trans} we show that the other transport coefficients can simply be expressed as ($\mat{L}_{ij}(\varepsilon)\equiv\mat{L}_{ij}(0,\varepsilon)$):
\begin{align}
\label{eq:appexplained2}
\mat{L}_{12}(\varepsilon)&=\varepsilon\mat{L}_{11}(\varepsilon)\ecom &
\mat{L}_{22}(\varepsilon)&=\varepsilon^2\mat{L}_{11}(\varepsilon)\ecom
\end{align}
and $\mat{L}_{12}(\varepsilon)=\mat{L}_{21}(\varepsilon)$. In this way, all the transport coefficients can be expressed in terms of the conductivity at zero temperature.

Close to zero temperature ($T\to0$) using the Sommerfeld expansion, we obtain:
\begin{subequations}
\label{eq:LlowT}
\begin{align}
\mat{L}_{11}(T,\mu)&\approx\mat{L}_{11}(\mu)+\frac{\pi^2}{6}T^2\partial^2_\mu\mat{L}_{11}(\mu)\ecom\\
\mat{L}_{12}(T,\mu)&\approx\mu\mat{L}_{11}(\mu)+\frac{\pi^2}{6}T^2\partial^2_\mu\left[\mu\mat{L}_{11}(\mu)\right]\ecom\\
\mat{L}_{22}(T,\mu)&\approx\mu^2\mat{L}_{11}(\mu)+\frac{\pi^2}{6}T^2\partial^2_\mu\left[\mu^2\mat{L}_{11}(\mu)\right]\edot
\end{align}
\end{subequations}
After changing to dimensionless units as in Eq. (\ref{eq:dimensionless}) the relevant parameter will be the dimensionless temperature $\mathcal{T}\defeq\ell_B T$. So the criteria for low temperatures is:
\begin{equation}
\frac{k_B T}{v\sqrt{\hbar eB}}\ll 1\ecom
\end{equation}
which shows that the temperature range where the low temperature approximation can be used increases with increasing magnetic fields. For $B=\SI{1}{\tesla}$ the criteria is $T\ll\SI{300}{\kelvin}$.

The experimentally measurable coefficients (conductivity ($\vb*{\sigma}$), Seebeck ($\vb{S}$) and thermal conductivity ($\vb*{\kappa}$) tensors) can be expressed using the transport coefficients as\cite{Mahan2000}:
\begin{subequations}
 \begin{align}
    \vb*{\sigma}&=e^2\vek{L}_{11}\ecom\\
    \vb{S}&=-\frac{1}{eT}\vek{L}_{11}^{-1}(\vek{L}_{12}-\mu\vek{L}_{11})\ecom\\
    \vb*{\kappa}&=\frac{1}{T}\left[\vek{L}_{22}-\vek{L}_{21}\vek{L}_{11}^{-1}\vek{L}_{12}\right]\edot
  \end{align}
\end{subequations}
Using Eq. (\ref{eq:LlowT}) at low temperatures :
\begin{subequations}
\begin{align}
\vb*{\sigma}(T,\mu)&\approx e^2\mat{L}_{11}(\mu)\ecom\\
\label{eq:seebeck}
\mat{S}(T,\mu)&\approx -\frac{\pi^2T}{3e}\mat{L}_{11}^{-1}(\mu)\partial_\mu\mat{L}_{11}(\mu)\ecom\\
\label{eq:WF}
\vb*{\kappa}(T,\mu)&\approx \frac{\pi^2}{3}T\mat{L}_{11}(\mu)\edot
\end{align}
\end{subequations}
As we can see the Mott's formula and the Wiedemann-Franz law hold.

\subsection{Hall conductivity $L_{xy}^{11}$}

For the Hall conductivity in Eq. (\ref{eq:L11full}) only the second part is needed since the product of matrix elements of the current operators (Eq. (\ref{eq:jmat})) are purely imaginary. Furthermore, since we are interested in the lowest order approximation in the impurities we can use the clean limit for $L_{xy}^{11}$:
\begin{widetext}
\begin{align}
\label{eq:Lxy11}
L^{11}_{xy}(\mu)&=\frac{1}{\pi V}\sum\limits_{\la,\lb} \Im{J^{(x)}_{\la\lb}J^{(y)}_{\lb\la}}\int\limits_{-\infty}^\mu \D\xi 2\partial_\xi \Re G_{\lb}^R(\xi)\Im G^R_{\la}(\xi)\bigg]\edot
\end{align}
With no impurities the imaginary part of the Green's function will be a Dirac delta $\Im G_{\la}^R(\xi) =-\pi\delta(\xi-E_{\la})$. With this the integral in Eq. (\ref{eq:Lxy11}) can be evaluated:
 \begin{equation}
        \label{eq:hall}
        L^{11}_{xy}(\mu) =-\frac{1}{2\pi^2\ell_B}\sum\limits_{n=0}^{\infty}\sum\limits_{\substack{\lambda,\lambda' = \pm 1 \\ s,s'=\pm1}} \int {\D \pdim_z} \left(U_{n\lambda s}^{n+1\lambda' s'}\right)^2 \frac{f(\lambda E_n-\mdim)-f(\lambda'E_{n+1}-\mdim)}{(\lambda E_n-\lambda' E_{n+1})^2}\ecom
      \end{equation}
\end{widetext}
where the summation over $s$ is taken only for $n\neq0$. As we have shown in Ref. \onlinecite{Konye2018} this can be expressed using the charge carrier density as (the sign in the previous paper is mistaken.): 
      \begin{equation}
	\label{eq:hallsimp}
        \sigma_{xy}=-\frac{en_e}{B}\edot
      \end{equation}
This result holds for finite temperatures and finite mass terms in arbitrary magnetic fields in the clean limit. In the quantum limit, we can express this using the chemical potential (Eq. (\ref{eq:chemlim})) as:
\begin{equation}
\label{eq:hallmu}
\sigma_{xy}=-\frac{e^2}{2\pi^2}\sqrt{\mu^2-\Delta^2}\edot
\end{equation}

\subsection{Conductivity $L_{xx}^{11}$}
From the matrix elements of the current operator (Eq. (\ref{eq:jx})) we can see that only the $n\neq n'$ Landau levels have a finite contribution. This means that without impurities the conductivity vanishes because the imaginary part of the Green's functions will be Dirac deltas centered around different energies.

As explained in Appendix \ref{app:vertex} the lowest order approximation of the conductivity in the case of impurities has two terms:
\begin{equation}
L^{11}_{xx}=L^{11(0)}_{xx}+L^{11(1)}_{xx}\ecom
\end{equation}
where $L^{11(1)}_{xx}$ is the vertex correction.

We start with calculating the first term:
\begin{align}
L^{11(0)}_{xx}(\mu)&=\frac{1}{\pi V}\sum\limits_{\la,\lb} \left\lvert J^{(x)}_{\la\lb}\right\rvert^2 \Im G^R_{\lb}(\mu) \Im G^R_{\la}(\mu)\ecom
\end{align}
where the impurity Green's function is taken from Eq. (\ref{eq:green}) and the matrix elements of the current operator are taken from Eq. (\ref{eq:jmat}). The Green's function and the scattering rate are redefined as $G(\varepsilon-\mu)\equiv G(\varepsilon)$ and $\Gamma(\varepsilon-\mu)\equiv\Gamma(\varepsilon)$. With the dimensionless units we obtain:
\begin{widetext}
\begin{equation}
        L_{xx}^{11(0)}(\mu) =\frac{1}{4\pi^3\ell_B}\sum\limits_{n=0}^\infty\sum\limits_{\substack{\lambda,\lambda' = \pm 1 \\ s=\pm1,s'=\pm1}} \int \D \pdim_z \left(U_{n\lambda s}^{n+1\lambda' s'}\right)^2\Im G^R_{n\lambda s}(\mdim) \Im G^R_{n+1\lambda's'}(\mdim)\ecom
\end{equation}
which is the same as the formula obtained in Ref. \onlinecite{Konye2018}. The imaginary part of the Green's function is expressed as:
 \begin{equation}  
   \Im{G_{\la}^R(\mdim)}=-\frac{\ell_B\Gamma_{\la}(\mdim,\pdim_z,B)}{(E_{\la}-\mdim)^2+(\ell_B\Gamma_{\la}(\mdim,\pdim_z,B))^2}\edot
 \end{equation}
In the quantum limit the scattering rate is small (if the impurity density is small enough) and the imaginary parts of the Green's function can be approximated as:
 \begin{align}
  \label{eq:greenlimit}  
   \Im{G_{0\lambda s}^R(\mdim)}&\approx-\pi\delta(E_{0\lambda s}-\mdim)\ecom &\Im{G_{n+1\lambda s}^R(\mdim)}&\approx-\frac{\ell_B\Gamma_{n+1\lambda s}(\mdim,\pdim_z,B)}{(E_{n+1\lambda s}-\mdim)^2}\edot
 \end{align}
In the lowest order of the impurity density we have to keep only the $n=0$ term which leads to:
\begin{align}
        L_{xx}^{11(0)} &=\frac{1}{4\pi^2}\sum\limits_{\substack{\lambda,\lambda' = \pm 1 \\ s'=\pm1}} \int \D \pdim_z  \left(U_{0\lambda \tilde{s}}^{1\lambda' s'}(\pdim_z,\ddim)\right)^2 \delta(E_{0\lambda s}-\mdim)\frac{\Gamma_{1\lambda' s'}(\mdim,\pdim_z,B)}{(E_{1\lambda's'}-\mdim)^2}\ecom
\end{align}
where $\tilde{s}=-\mathrm{sgn}(\pdim_z)$. After performing the $\pdim_z$ integral and using the Eq. (\ref{eq:chemlim}) form of the chemical potential at high fields (assuming $\mu>0$) we obtain:
\begin{align}
	\label{eq:Lxxfull}
        L_{xx}^{11(0)} &=\frac{1}{4\pi^2}\sum\limits_{\substack{\lambda' = \pm 1 \\ s'=\pm1,\xi=\pm1}}  \left(U_{01 \tilde{s}}^{1\lambda' s'}(\xi\ddim_B,\ddim)\right)^2 \frac{\sqrt{\ddim_B^2+\ddim^2}}{\ddim_B}\frac{\Gamma_{1\lambda' s'}\left(\sqrt{\ddim_B^2+\ddim^2},\xi\ddim_B\right)}{\left(\lambda'\sqrt{2+\ddim_B^2+\ddim^2}-\sqrt{\ddim_B^2+\ddim^2}\right)^2}\edot
\end{align}
\end{widetext}
In the very high field limit we can use Eq. (\ref{eq:gambigB}) for the scattering rate:
\begin{equation}
	\label{eq:Lxx0as}
        L_{xx}^{11(0)} =\begin{cases}
       n_iu_i^2\ell_B^2I &\Delta_B\gg\Delta\\
        n_iu_i^2\ell_B^2J &\Delta_B\ll \Delta\\
     \end{cases}\propto\frac{1}{B}\ecom
\end{equation}
where:
\begin{subequations}
\begin{align}
I &\defeq \sum\limits_{\substack{\lambda' = \pm 1 \\ s'=\pm1,\xi=\pm1}} \frac{1}{8\pi^2} \left(U_{01 -\xi}^{1\lambda' s'}(0,0)\right)^2 I_{1\lambda' s'}\ecom\\
J &\defeq \sum\limits_{\substack{\lambda' = \pm 1 \\ s'=\pm1,\xi=\pm1}} \frac{1}{8\pi^2}\left(U_{01 -\xi}^{1\lambda' s'}(0,0)\right)^2 J_{1\lambda' s'}\edot 
\end{align}
\end{subequations}
The conductivity ($\sigma^0_{xx}=e^2L^{11(0)}_{xx}$) calculated numerically using Eq. (\ref{eq:Lxxfull}) can be seen in Fig. \ref{fig:Lxx}.	
\plot{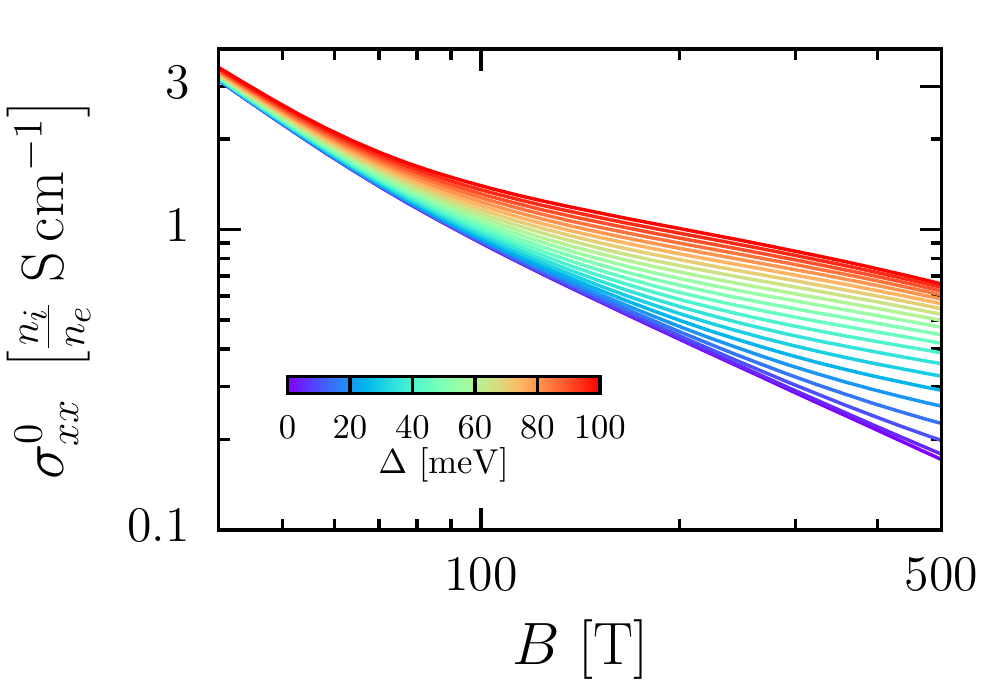}{width=.45\textwidth}{The conductivity without the vertex correction in the quantum limit (\ref{eq:Lxxfull}) as a function of the magnetic field using different mass terms. The parameters used are: $n_e=10^{18}\si{\per\cubic\centi\metre}$ and $v=10^6\si{\metre\per\second}$.}

As we can see in the high field limit we recover the asymptotic behavior described in Eq. (\ref{eq:Lxx0as}), and at lower fields the effect of the mass term becomes less relevant. The real asymptotic behavior is only reached at very high fields. In the intermediate region where the curves with different mass term start to diverge from each other the decrease is weaker then $B^{-1}$. As a consequence in the high field limit the quantitative value of the conductivity is larger for larger values of the mass term.

Now we move on with the calculation of the vertex correction, $L^{11(1)}_{xx}$. As explained in Appendix \ref{app:vertex} in the lowest order approximation this can be calculated as:
\begin{widetext}
\begin{subequations}
\begin{align}
L_{xx}^{11(1)}(\mu)&=\frac{1}{\pi V}\sum\limits_{\ln,\ln',\lm,\lm'} V_{\ln\lm}^{\ln'\lm'}{J}^{(x)}_{\ln\ln'}{J}^{(x)}_{\lm'\lm}\Im C_{\ln\lm}^{\ln'\lm'}(\mu)\ecom\\
V_{\ln\lm}^{\ln'\lm'}&\defeq\frac{1}{V^3}\sum\limits_{\substack{\vek{k},\vek{k'}\\ \vek{q}}}n_i|u_{\vek{q}}|^2 \gvek{\phi}_{\ln'}^\dag(\vek{k'})\gvek{\phi}_{\lm'}(\vek{k'-q})\gvek{\phi}_{\lm}^\dag(\vek{k-q})\gvek{\phi}_{\ln}(\vek{k})\ecom\\
\Im C_{\ln\lm}^{\ln'\lm'}&=\frac{[(E_{\ln}-\mu)\Gamma_{\lm} +(E_{\lm}-\mu)\Gamma_{\ln}][ (E_{\lm'}-\mu)\Gamma_{\ln'}+(E_{\ln'}-\mu)\Gamma_{\lm'}]}{\Gamma_{\ln}\Gamma_{\lm}\Gamma_{\lm'}\Gamma_{\ln'}}\Im G_{\ln}^R \Im G_{\lm}^R \Im G_{\lm'}^R \Im G_{\ln'}^R\ecom
\end{align}
\end{subequations}
where the Green's functions and scattering rates are evaluated at the chemical potential ($\Gamma_{\la}\equiv \Gamma_{\la}(\mu,p_z,B)$ and $G^R_{\la}\equiv G^R_{\la}(\mu)$). For the quantum numbers we use: $\ln\equiv(n,\lambda,s,p_z,p_y)$, $\ln'\equiv(n',\lambda',s',p_z,p_y)$, $\lm\equiv(m,\gamma,t,p_z',p_y')$ and $\lm'\equiv(m',\gamma',t',p_z',p_y')$. The part with the impurity potential can be expressed similarly to the scattering rate:
\begin{align}
V_{n\lambda s;m\gamma t}^{n'\lambda's';m'\gamma't'}(\pdim_y,\pdim_y',\pdim_z,\pdim_z') &=\frac{n_i}{\ell_B^3} \int\frac{\D^3 \qdim}{(2\pi)^3}u_{\vek{\qdim}}^2 \delta_{\pdim_y-\qdim_y,\pdim_y'}\delta_{\pdim_z-\qdim_z,\pdim_z'} F_{n'\lambda's'}^{m'\gamma' t'}(\vb*{\qdim},\pdim_z)F_{n\lambda s}^{m\gamma t*}(\vb*{\qdim},\pdim_z)\ecom
\end{align}
where $F$ is defined in Eq. (\ref{eq:four}). In the lowest order approximation in the impurity density we can again use Eq. (\ref{eq:greenlimit}) for the imaginary part of the Green's function. Using the matrix elements of the current operator keeping only the lowest Landau indexes (higher indexes will have little contribution at high fields), we obtain: 
\begin{equation}
L^{11(1)}_{xx}=\frac{2\pi\ell_B^4}{V}\hspace{-10pt}\sum\limits_{\substack{\lambda,\lambda',\gamma,\gamma'\\s',t\\\pdim_z,\pdim_z',\pdim_y,\pdim_y'}} \hspace{-10pt}V_{0\lambda \tilde{s};1\gamma t}^{1\lambda's';0\gamma'\tilde{t}'}(\pdim_y,\pdim_y',\pdim_z,\pdim_z') U_{0\lambda \tilde{s}}^{1\lambda's'}(\pdim_z)U_{0\gamma' \tilde{t}'}^{1\gamma t}(\pdim_z')\frac{\delta(E_{0\gamma'\tilde{t}'}(\pdim_z')-\mdim)}{E_{1\gamma t}(\pdim_z')-\mdim}\frac{\delta(E_{0\lambda \tilde{s}}(\pdim_z)-\mdim)}{E_{1\lambda' s'}(\pdim_z)-\mdim}\ecom
\end{equation}
where we use the fact that $V$ is real and the symmetry properties in the quantum numbers. Using the Kronecker deltas in $V$ and evaluating all momentum integrals except $\qdim_x$ and $\qdim_y$, we obtain:
\begin{subequations}
\begin{align}
\label{eq:vertexfinal}
        L_{xx}^{11(1)} =\frac{n_iu_i^2\ell_B^2}{4\pi^2}\sum\limits_{\substack{\lambda',\gamma=\pm 1\\ s',t=\pm1\\\xi,\eta=\pm1}}  \frac{U_{01\tilde{s}}^{1\lambda's'}(\xi\ddim_B,\ddim)U_{01\tilde{t}'}^{1\gamma t}(\eta\ddim_B,\ddim) \frac{\sqrt{\ddim_B^2+\ddim^2}}{\ddim_B}\Theta_{01 \tilde{s};1\gamma t}^{1\lambda's';01\tilde{t}'}\left((\xi+\eta)\ddim_B,\xi\ddim_B\right)}{\left(\lambda'\sqrt{2+\ddim_B^2+\ddim^2}-\sqrt{\ddim_B^2+\ddim^2}\right)\left(\gamma\sqrt{2+\ddim_B^2+\ddim^2}-\sqrt{\ddim_B^2+\ddim^2}\right)}\\
\Theta_{n\lambda s;m\gamma t}^{n'\lambda's';m'\gamma't'}(\qdim_z,\pdim_z)\defeq\int\frac{\D \qdim_x\D \qdim_y}{(2\pi)^2} \frac{ \frac{\sqrt{\ddim_B^2+\ddim^2}}{\ddim_B} F_{n'\lambda's'}^{m'\gamma' t'}(\vb*{\qdim},\pdim_z)F_{n\lambda s}^{m\gamma t*}(\vb*{\qdim},\pdim_z)}{\left(\bm{\qdim}^2+\frac{u_e}{2\pi^2}\frac{\sqrt{\ddim_B^2+\ddim^2}}{\ddim_B}\right)^2}\edot
\end{align}
\end{subequations}
The structure of the vertex correction is very similar to that of $L_{xx}^{11(0)}$, but the origin of each term is different. Similarly to how we calculated the limit of the scattering rate and $L_{xx}^{11(0)}$ we can evaluate the high field limit of the vertex correction. Since the structure is similar we get the same asymptotic behavior as in Eq. (\ref{eq:Lxx0as}) but the proportionality constants will be different:
\begin{equation}
\label{eq:vertexasy}
        L_{xx}^{11(1)} =\begin{cases}
       n_iu_i^2\ell_B^2\tilde{I} &\Delta_B\gg\Delta\\
        n_iu_i^2\ell_B^2\tilde{J} &\Delta_B\ll \Delta\\
     \end{cases}\propto\frac{1}{B}\ecom
\end{equation}
where:
\begin{subequations}
\begin{align}
\tilde{I} &\defeq \frac{1}{8\pi^2}\sum\limits_{\substack{\lambda',\gamma=\pm 1\\ s',t=\pm1\\\xi,\eta=\pm1}}  \frac{U_{01\tilde{s}}^{1\lambda's'}(0,0)U_{01\tilde{t}'}^{1\gamma t}(0,0)}{\lambda'\gamma}\int\frac{\D \qdim_x\D \qdim_y}{(2\pi)^2} \frac{ F_{1\lambda's'}^{01\tilde{t}'}(\qdim_x,\qdim_y,\qdim_z=0,0)F_{01\tilde{s}}^{1\gamma t*}(\qdim_x,\qdim_y,\qdim_z=0,0)}{\left(\qdim_x^2+\qdim_y^2+\frac{u_e}{2\pi^2}\right)^2}\ecom\\
\tilde{J} &\defeq \frac{1}{8\pi^2}\sum\limits_{\substack{\lambda',\gamma=\pm 1\\ s',t=\pm1\\\xi,\eta=\pm1}}  \frac{U_{01\tilde{s}}^{1\lambda's'}(0,0)U_{01\tilde{t}'}^{1\gamma t}(0,0)}{\lambda'\gamma}\int\frac{\D \qdim_x\D \qdim_y}{(2\pi)^2} \frac{ F_{1\lambda's'}^{01\tilde{t}'}(\qdim_x,\qdim_y,\qdim_z=0,0)F_{01\tilde{s}}^{1\gamma t*}(\qdim_x,\qdim_y,\qdim_z=0,0)}{\left(\frac{u_e}{2\pi^2}\right)^2}\edot
\end{align}
\end{subequations}
\end{widetext}
The vertex correction contribution was numerically calculated from Eq. (\ref{eq:vertexfinal}). The results can be seen on Fig. \ref{fig:Lxx_vertex}.
\plot{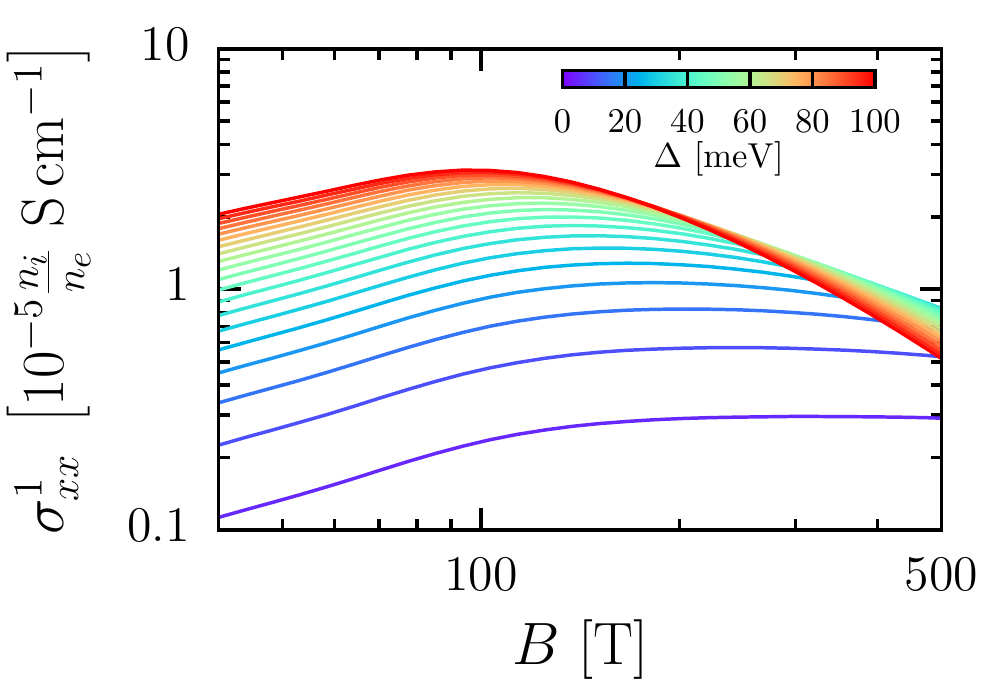}{width=.45\textwidth}{The vertex correction of the conductivity in the quantum limit (\ref{eq:vertexfinal}) as a function of the magnetic field using different mass terms. The parameters used are: $n_e=10^{18}\si{\per\cubic\centi\metre}$ and $v=10^6\si{\metre\per\second}$.}

At high fields we do not recover the asymptotic behavior described in Eq. (\ref{eq:vertexasy}), but instead we get a more rapid decrease. This is because in the analytic formula we assumed that $\tilde{I}$ and $\tilde{J}$ are non-zero. From the numerics we can see that they are numerically zero. This gives an extra decrease when $\ddim_B,~\ddim\to 0$. From the numerical results we can see that the vertex correction is several orders of magnitude smaller than $\sigma_{xx}^{0}$. This is because in the summation over indices in Eq. (\ref{eq:vertexfinal}) $\Theta$ is close to zero when the matrix elements of the current operators give finite values. Furthermore, the vertex correction at $\Delta=0$ is numerically zero (it is not visible on the figure). When $\ddim\gg0$ the vertex correction becomes more relevant, but at reasonable values of the mass term it is negligible. In the following section we will neglect the vertex correction, since it has a very small contribution to the overall conductivity.

  \subsection{Seebeck tensor}

Since the thermal conductivity is proportional to $T\vb*{\sigma}(T,\mu)$ as in Eq. (\ref{eq:WF}), we only focus on the Seebeck tensor.

To calculate the Seebeck tensor (\ref{eq:seebeck}) we need the resistivity tensor and the derivative of the conductivity with respect to the chemical potential. The independent elements of the Seebeck tensor are:
\begin{subequations}
\begin{align}
S_{xx} &= -T\frac{\pi^2}{3e}\frac{\sigma_{xx}\partial_\mu\sigma_{xx}+\sigma_{xy}\partial_\mu\sigma_{xy}}{\sigma_{xx}^2+\sigma_{xy}^2}\ecom\\
S_{xy} &= -T\frac{\pi^2}{3e}\frac{\sigma_{xx}\partial_\mu\sigma_{xy}-\sigma_{xy}\partial_\mu\sigma_{xx}}{\sigma_{xx}^2+\sigma_{xy}^2}\edot
\end{align}
\end{subequations}

The result for the Seebeck ($S_{xx}$) and Nernst ($S_{xy}$) coefficients depends strongly on the Hall angle ($\tan{\vartheta_H}=\sigma_{xy}/\sigma_{xx}$), which is mainly determined by the ratio of charge carrier density and the impurity density. Based on experimental results for the Hall angle\cite{Leahy2018} we will assume $\sigma_{xy}>\sigma_{xx}$. This is in good agreement with the assumption that the impurity density is not too high. For simplicity from now on we will assume that $\tan{\vartheta_H}\gg 1$. In this case the elements of the Seebeck tensor are:
\begin{subequations}
\begin{align}
S_{xx} &= -T\frac{\pi^2}{3e}\frac{\partial_\mu\sigma_{xy}}{\sigma_{xy}}\ecom\\
S_{xy} &= -T\frac{\pi^2}{3e}\frac{\sigma_{xx}\partial_\mu\sigma_{xy}-\sigma_{xy}\partial_\mu\sigma_{xx}}{\sigma_{xy}^2}\edot
\end{align}
\end{subequations}

In the quantum limit the derivative of the Hall conductivity (\ref{eq:hallmu}) with respect to the chemical potential is:
\begin{equation}
\partial_\mu\sigma_{xy}=-\frac{e^2}{2\pi^2}\frac{\sqrt{\Delta_B^2+\Delta^2}}{\Delta_B}\edot
\end{equation}
With this the Seebeck coefficient can be expressed analytically as:
\begin{equation}
  \label{eq:Sxx}
        S_{xx} =-T\frac{\pi^2}{3e}\frac{\sqrt{\Delta_B^2+\Delta^2}}{\Delta_B^2}\propto\begin{cases}
       B &\Delta_B\gg\Delta\\
       B^2 &\Delta_B\ll \Delta\\
     \end{cases}\edot
\end{equation}
For the massless case we recover the linear non-saturating result obtained experimentally and theoretically in Refs. \onlinecite{Skinner2018,Liang2013}. For the massive case we get a significantly different behavior since the Seebeck coefficient is proportional to the square of the magnetic field. As we can see this result does not depend on the impurity density as long as the impurities can be neglected in the calculation of the Hall-conductivity. The Seebeck coefficient (\ref{eq:Sxx}) using different mass terms can be seen in Fig. \ref{fig:Sxx}.
\plot{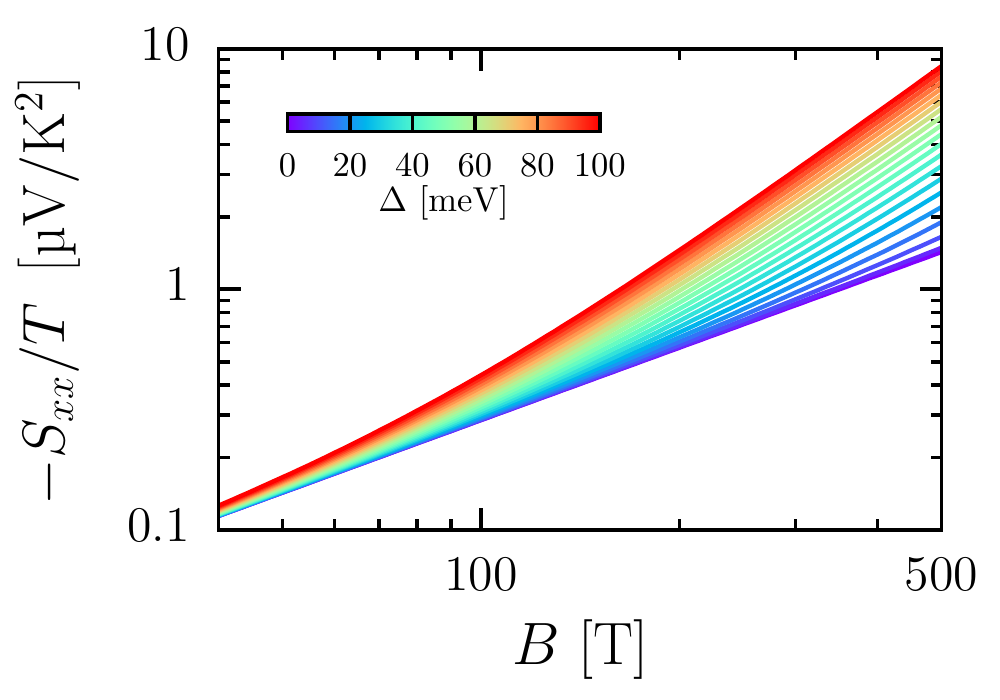}{width=.45\textwidth}{The Seebeck coefficient in the quantum limit (\ref{eq:Sxx}) as a function of the magnetic field using different mass terms. The parameters used are: $n_e=10^{18}\si{\per\cubic\centi\metre}$ and $v=10^6\si{\metre\per\second}$.}

\begin{figure}
      \includegraphics[width=.45\textwidth]{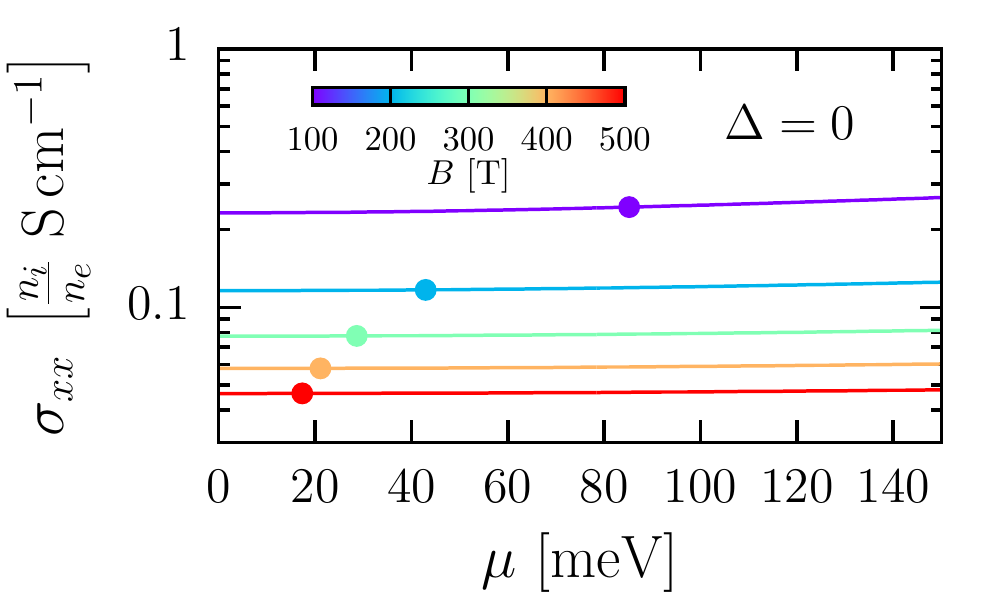}
      \vspace{-20pt}
      \includegraphics[width=.45\textwidth]{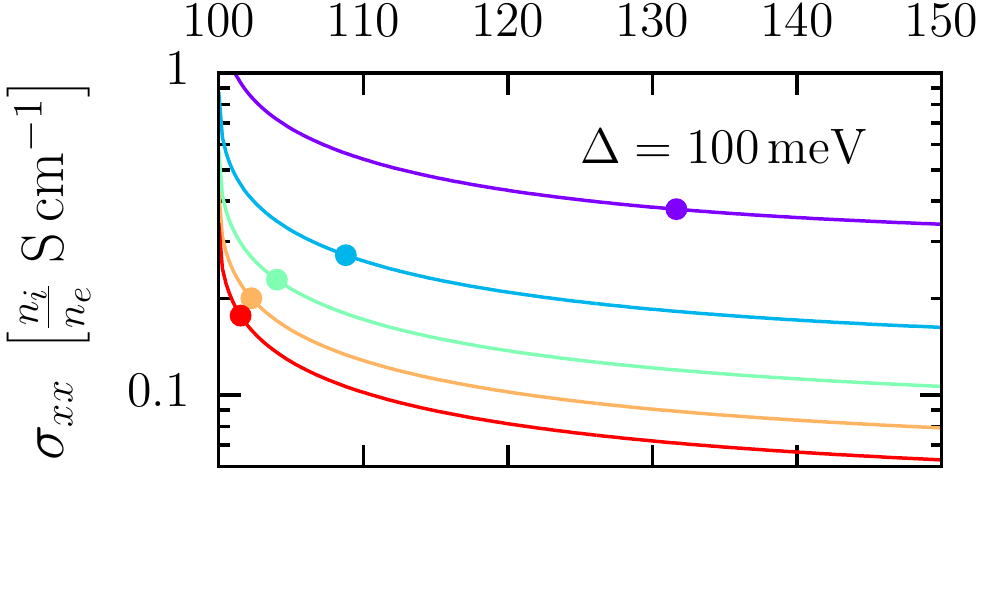}
      \caption{\label{fig:Lxxmu}The conductivity in the quantum limit (\ref{eq:Lxxfull}) as a function of the chemical potential at different magnetic fields. The mass term is $\Delta=0$ (top plot) and $\Delta=\SI{100}{\milli\electronvolt}$ (bottom plot). The solid circles show the values of the chemical potential when $n_e=10^{18}\si{\per\cubic\centi\metre}$ at the given magnetic fields, which is the point where the derivative in Eq. (\ref{eq:Sxycom}) has to be evaluated.}
 \end{figure}

\begin{figure}
      \includegraphics[width=.45\textwidth]{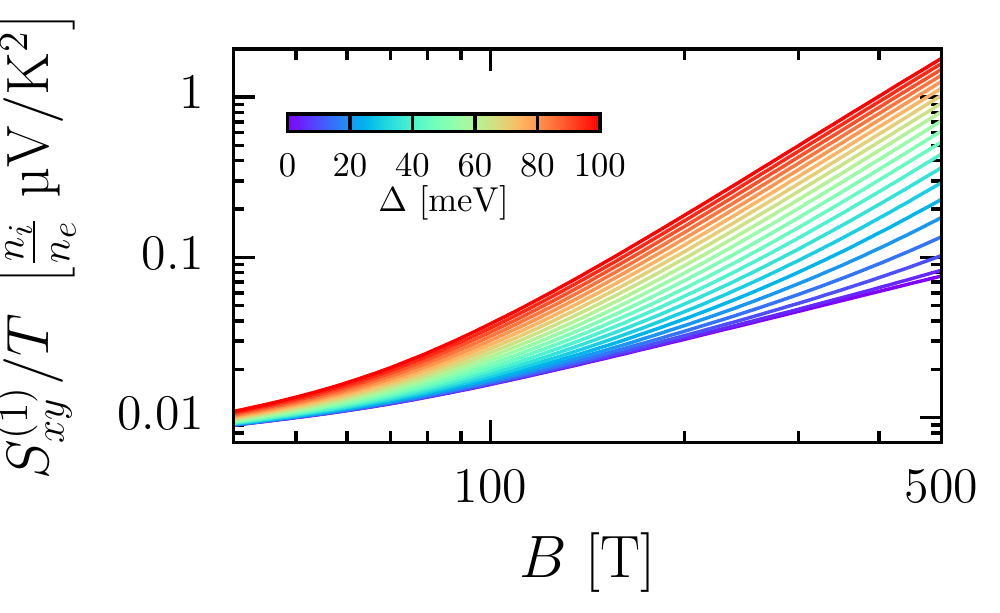}
      \vspace{-20pt}
      \includegraphics[width=.45\textwidth]{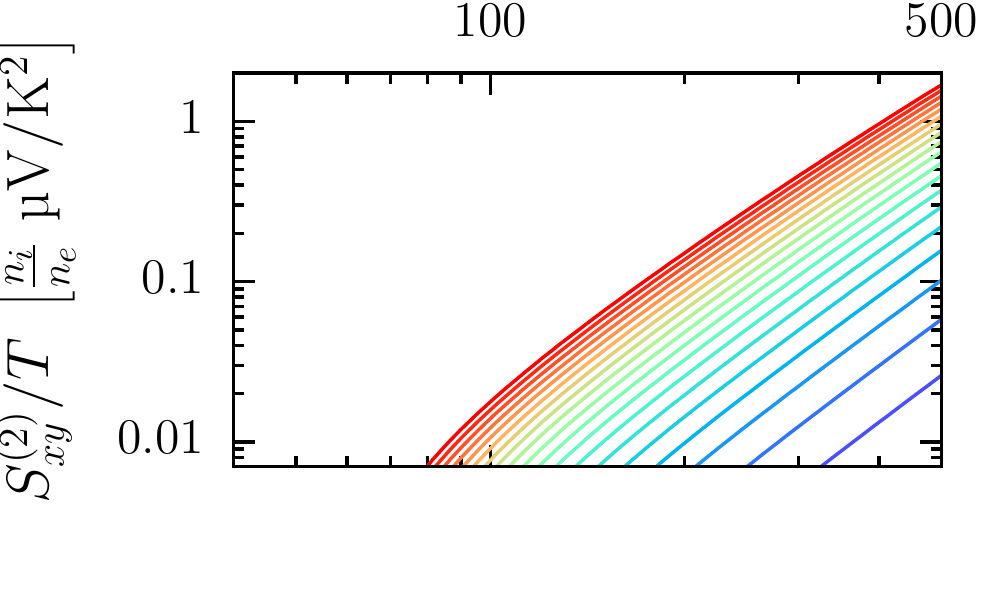}
      \caption{\label{fig:Sxy1}The two components of the Nernst coefficient in the quantum limit (\ref{eq:Sxycom}) as a function of the magnetic field using different mass terms. The parameters used are: $n_e=10^{18}\si{\per\cubic\centi\metre}$ and $v=\SI{1e6}{\metre\per\second}$}
 \end{figure}

The Nernst-coefficient can be divided in two terms as:
\begin{equation}
\label{eq:Sxycom}
S_{xy}=\underbrace{S_{xx}\frac{\sigma_{xx}}{\sigma_{xy}}}_{S_{xy}^{(1)}}+\underbrace{T\frac{\pi^2}{3e}\frac{\partial_\mu\sigma_{xx}}{\sigma_{xy}}}_{S_{xy}^{(2)}}\edot
\end{equation}
As we showed in previous sections $\sigma_{xx}\propto\sigma_{xy}$ at high fields, this means that the magnetic field dependence of $S_{xy}^{(1)}$ is qualitatively the same as $S_{xx}$. For the second term we need to evaluate $\partial_\mu\sigma_{xx}$. Since the chemical potential dependence affects many components of the conductivity through $\Delta_B$ in Eq. (\ref{eq:Lxxfull}), we calculate the derivative numerically. The conductivity as a function of the chemical potential can be seen in Fig. \ref{fig:Lxxmu}.
In the massless case the conductivity is almost independent of the chemical potential, but in the massive case we can see that the dependence becomes very strong close to the bottom of the Landau level, caused by the diverging density of states. Using the numerical derivative for the conductivity we can calculate $S_{xy}^{(2)}$. The results for the two components of the Nernst coefficient can be seen on Fig. \ref{fig:Sxy1}.

The first component is very similar to the Seebeck coefficient, but it is suppressed by $\cot{\vartheta_H}$. The second component is very small (at certain regions even negative) at low fields and low mass terms. At higher fields in the case of finite mass term it has the same dependence as $S_{xy}^{(1)}$ and roughly the same value.

The total Nernst coefficient can be seen in Fig. \ref{fig:Sxy}.
\plot{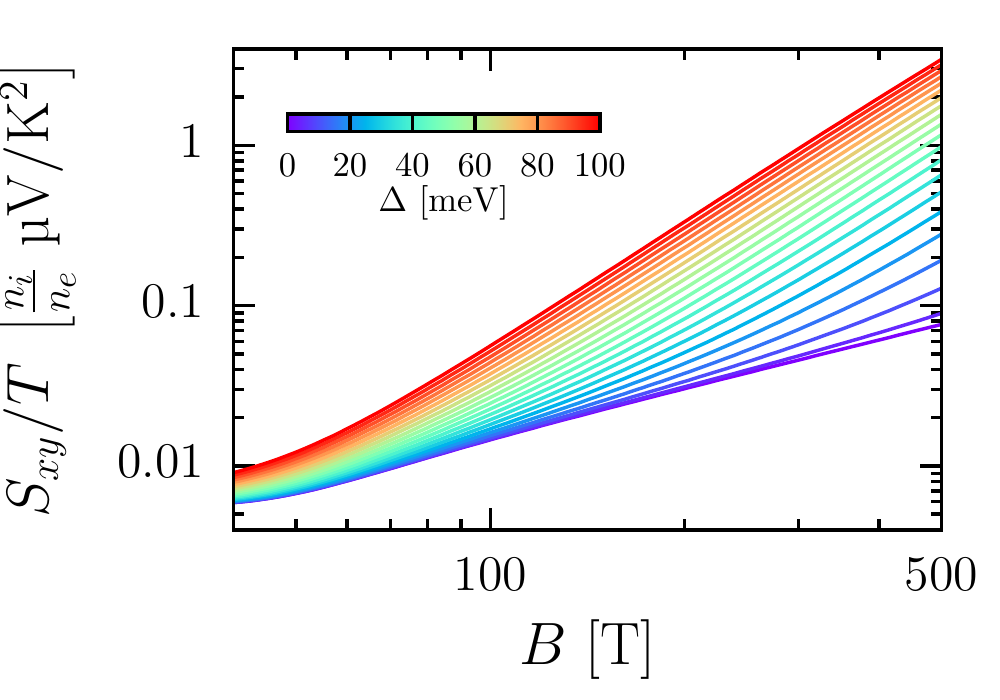}{width=.45\textwidth}{The Nernst coefficient in the quantum limit (\ref{eq:Sxycom}) as a function of the magnetic field using different mass terms. The parameters used are: $n_e=10^{18}\si{\per\cubic\centi\metre}$ and $v=10^6\si{\metre\per\second}$.}
The dependence is very similar to that of the Seebeck coefficient. The main difference is at lower fields where $\sigma_{xx}$ differs from the asymptotic behavior.

\section{Summary and Discussions}

We studied the massive Dirac Hamiltonian in a constant magnetic field. We show the analytic asymptotic behaviors of the chemical potential the screening wavenumber and the scattering rate based on the results obtained in Ref. \onlinecite{Konye2018}. The important energy scale is $\Delta_B=2\pi^2 n_e\ell_B^2$ and the mass term $\Delta$ becomes relevant when $\Delta_B<\Delta$. For realistic systems this happens at high fields which are hard to realize experimentally. At lower fields the effect of the mass term is negligible. The asymptotic behavior of these quantities at high magnetic fields is:
\begin{subequations}
\begin{align}
\mu&\propto B^{-1} &\kappa^2&\propto B & \Gamma&\propto B^{-1}& \Delta_B&\gg\Delta\ecom\\
\mu&\sim\Delta&\kappa^2 &\propto B^2 & \Gamma&\propto B^{-2} & \Delta_B&\ll\Delta\edot
\end{align}
\end{subequations}

In Sec. \ref{sec:transport} we study the thermoelectric transport coefficients based on the formalism developed by Luttinger\cite{Luttinger1964,Smrcka1977}. We prove the validity of the Sommerfeld-Bethe relations\cite{OgataFukuyama} and as a consequence the Mott formula and the Wiedemann-Franz law. We show that the temperature range where the low temperature approximation can be applied expands with increasing magnetic fields.

First, we calculated the electric conductivity tensor. We showed that the Hall conductivity in the clean limit is inversely proportional to the magnetic field, as in usual systems. In the case of $\sigma_{xx}$ impurities are necessary to get a non-zero result. In our previous study\cite{Konye2018} we have shown that the magnetic field dependence of the scattering rate directly affects the magnetic field dependence of the conductivity. The screened charged impurities are necessary to reproduce the $B^{-1}$ dependence of the conductivity and thus the linear magnetoresistance, consistently with previous studies for the $2\times2$ Weyl Hamiltonian\cite{Abrikosov1998,Xiao2017,Klier2017a}. We calculated the vertex correction and showed that it is negligible for realistic parameter regimes, which justifies the assumptions in previous studies\cite{Abrikosov1998,Xiao2017,Konye2018}. In the case of finite mass term we showed that the high field limit is qualitatively the same as in the massless case i.e. proportional to $B^{-1}$ but with different numerical prefactors. At intermediate magnetic fields we find that the conductivity decreases at a slower rate than $B^{-1}$. At high fields the scattering rate decreases at a higher rate in the massive case, but the density of states becomes larger close to the bottom of the Landau level. These two effects compensate each other causing the same behavior as in the massless case.

Then, using the Mott-formula we studied the Seebeck and Nernst coefficients. We assumed that the Hall conductivity is larger than the diagonal conductivity, and thus the Hall-angle is large. This is a reasonable assumption if the impurity density is low and it is consistent with experimental results\cite{Leahy2018}. With this $S_{xx}$ is linear in the massless case consistently with Ref. \onlinecite{Skinner2018}, but in the massive case it increases quadratically. This is also consistent with the experimental result for massive Dirac electrons\cite{Liang2013} where they found the thermopower to be linear, since the magnetic field is not high enough in the experiment to see the different behavior. While the qualitative difference does not appear in the conductivity we see it in the Seebeck coefficient but only at very high fields. As we saw the Nernst coefficient behaves very similarly to the Seebeck coefficient. This is mainly caused by the fact that the Hall angle saturates at high fields which makes $S_{xy}\propto S_{xx}$.

Summarizing the asymptotic results for the magnetic field and temperature dependence of the measurable quantities close to zero temperature and high magnetic fields we got:
\begin{subequations}
\begin{align}
\sigma_{xx}&\propto \frac{n_i}{B} &S_{xx}&\propto \frac{TB}{n_e} & \Delta_B&\gg\Delta\ecom\\
\sigma_{xx}&\propto \frac{n_i}{B} &S_{xx}&\propto \frac{TB^2\Delta}{n_e^2} & \Delta_B&\ll\Delta\ecom\\
\sigma_{xy}&\propto \frac{n_e}{B} &S_{xy}&\propto \frac{n_iTB}{n_e^2} & \Delta_B&\gg\Delta\ecom\\
\sigma_{xy}&\propto \frac{n_e}{B} &S_{xx}&\propto \frac{n_iTB^2\Delta}{n_e^3} &\Delta_B&\ll\Delta\edot
\end{align}
\end{subequations}

Since the Wiedemann-Franz law holds as in Eq. (\ref{eq:WF}), the thermal conductivity is simply proportional to the electric conductivity. To get a complete picture for the thermal conductivity the contribution from phonons would also be necessary.

\begin{acknowledgments}
We thank H. Matsuura, and H. Maebashi for useful discussions. This work was supported by a Grant-in-Aid for Scientific Research (B) on \lq\lq Multiband effects in magnetic responses and transport properties'' (No. 18H01162) from the MEXT of the Japanese Government.
\end{acknowledgments}

\appendix
\section{Current operators of a general multi-band Hamiltonian}
\label{app:curr}
In this appendix, we derive the current operators of a general multi-band Hamiltonian without interactions based on the continuum equation. An alternative method is given in Ref. \onlinecite{OgataFukuyama} for a single-band Hamiltonian with electron-phonon and finite-range mutual interactions.
We assume that the many-body Hamiltonian can be written as:
\begin{equation}
\mathcal{H} = \sum\limits_{a,b}\int\D^3 x~ \Psi_a^\dag(\vek{x}) H_{ab}(\vek{p}+e\vek{A}(\vek{x}),\vek{x}) \Psi_b(\vek{x})\ecom
\end{equation}
where $\vek{H}$ is an arbitrary Hermitian matrix that is a function of the momentum and coordinate. $\vek{A}$ is an arbitrary vector potential describing the magnetic field. From now on the summation over $a$ and $b$ will not be explicitly written, but the Einstein summation convention is used. In the coordinate representation the momentum becomes a differential operator which can be described with an appropriate singular kernel, thus the Hamiltonian can be written as:
\begin{equation}
\label{eq:kernel}
\mathcal{H} = \int\D^3 x\D^3 x'~ \Psi_a^\dag(\vek{x}) H_{ab}(\vek{x},\vek{x}') \Psi_b(\vek{x'})\edot
\end{equation}
The particle current and the energy current can be expressed through the continuity equation:
\begin{align}
\label{eq:contin}
\partial_t \varrho + \mathrm{div}\vek{j} &= 0\ecom&
\partial_t h + \mathrm{div}\vek{j}_E &= 0\edot
\end{align}
The particle and energy density operators are defined to be Hermitian as:
\begin{subequations}
\begin{align}
\varrho(x)&=\gvek{\Psi}^\dag(\vek{x})\gvek{\Psi}(\vek{x})\ecom\\
h(x)&=\int \D^3x' \frac{1}{2}\left[\gvek{\Psi}^\dag(\vek{x})\vek{H}(\vek{x},\vek{x}') \gvek{\Psi}(\vek{x}')+(x\leftrightarrow x') \right]\edot
\end{align}
\end{subequations}
From now on $\vek{j}_1\equiv\vek{j}$, $\vek{j}_2\equiv\vek{j}_E$, $\varrho_1\equiv\varrho$ and $\varrho_2\equiv h$. The many body current operators ($\rvek{J}_i$) can be expressed with the density operator using Eq. (\ref{eq:contin}) as:
\begin{align}
\nonumber
\mathcal{J}^{(i)}_\alpha  &\defeq \int \D^3 x j^{(i)}_\alpha(\vek{x}) = \int \D^3 x \sum\limits_\beta (\partial_\beta x_\alpha) j^{(i)}_\beta(\vek{x})\\
& = -\int \D^3 x \sum\limits_\beta x_\alpha (\partial_\beta j^{(i)}_\beta(\vek{x})) = \int \D^3 x~ x_\alpha \partial_t \varrho_i(\vek{x}) \ecom
\end{align}
where we neglected surface terms during partial integration. Introducing the polarization operators defined as:
\begin{equation}
\va{\rvek{P}}_i \defeq \int \D^3 x~ \va{x} \varrho_i(\vek{x})\ecom
\end{equation}
the current operators are:
\begin{equation}
\va{\rvek{J}}_i = \partial_t \va{\rvek{P}}_i = i[\mathcal{H},\va{\rvek{P}}_i]\edot
\end{equation}
Using the previous definitions and the anticommutator of fermionic field operators:
\begin{equation}
\{ \Psi_a(\vek{x}),\Psi_b^\dag(\vek{x}')\} =\delta_{ab}\delta(x-x')\ecom
\end{equation}
the current operators become:
\begin{subequations}
\begin{align}
\label{eq:currker}
\va{\rvek{J}}_1 = i\int\D^3x\D^3x'& \Psi_a^\dag(\vek{x}) H_{ab}(\vek{x},\vek{x}')(\va{x}'-\va{x}) \Psi_b(\vek{x}')\ecom\\
\nonumber
\va{\rvek{J}}_2 = \frac{i}{2}\int\D^3x\D^3x'&\D^3x'' \Psi_a^\dag(\vek{x})H_{ac}(\vek{x},\vek{x}'')\times\\
\label{eq:qcurrker}
&\times(\va{x}'-\va{x}) H_{cb}(\vek{x}'',\vek{x}')\Psi_b(\vek{x}')\edot
\end{align}
\end{subequations}
Using differential operators these can be expressed as:
\begin{equation}
\label{eq:currdiff}
\va{\rvek{J}}_i = \int\D^3x \Psi_a^\dag(\vek{x},t) \vec{J}^{(i)}_{ab}(\vek{p},\vek{x}) \Psi_b(\vek{x},t)\ecom\\
\end{equation}
where (using $[f(\vek{p}),\vek{x}]=-i\grad_{\vek{p}}f(\vek{p})$):
\begin{subequations}
\begin{align}
\vec{J}^{(1)}_{ab}(\vek{p},\vek{x})&=\grad_{\vek{p}}H_{ab}(\vek{p}+e\vek{A}(\vek{x}),\vek{x})\ecom\\
\vec{J}^{(2)}_{ab}(\vek{p},\vek{x})&=\frac{1}{2}[\vec{J}^{(1)}_{ac}H_{cb}+H_{ac}\vec{J}^{(1)}_{cb}]\edot
\end{align}
\end{subequations}

If the Hamiltonian includes impurities in the form of $V(\vek{x})$ we can see that it doesn't affect the particle current, but it appears in the energy current. Thus, in order to calculate the energy current the matrix elements of the impurity potential would be necessary. This can be avoided by expressing the energy current with the current operator. A similar argument for a single-band Hamiltonian can be found in Refs. \onlinecite{Jonson1990,OgataFukuyama}. We start by defining:
\begin{equation}
  \label{eq:manybodyJ}
  \va{\rvek{J}}_1(\tau,\tau') \defeq \int\D^3x\gvek{\Psi}^\dag(\vek{x},\tau)\va{J}_1\gvek{\Psi}(\vek{x},\tau')\ecom
\end{equation}
where we use the $\tau$ imaginary times in the Matsubara formalism. With this the many-body current operator is:
\begin{equation}
  \va{\rvek{J}}_1(\tau) = \lim\limits_{\tau'\to\tau^-} \va{\rvek{J}}_1(\tau,\tau')\edot
\end{equation}
Using the grand canonical Hamiltonian ($\mathcal{K}=\mathcal{H}-\mu\mathcal{N}$) the $\tau$ derivative of an arbitrary $A$ operator is:
\begin{equation}
  \partial_\tau A(\tau) = [\mathcal{K},A(\tau)]\edot
\end{equation}
It can be shown that:
\begin{subequations}
\begin{align}
  \partial_\tau {\Psi}^\dag_a(\vek{x},\tau) &= \int\D^3x' {\Psi}^\dag_b(\vek{x}',\tau)K_{ba}(\vek{x}',\vek{x})\ecom\\
  \partial_\tau {\Psi}_a(\vek{x},\tau) &= -\int\D^3x' K_{ab}(\vek{x}',\vek{x}){\Psi}_b(\vek{x}',\tau)\edot
\end{align}
\end{subequations}
With these the energy current can be expressed as:
\begin{equation}
  \label{eq:currmahan}
  \va{\rvek{J}}_2(\tau) =\lim\limits_{\tau'\to\tau^-}\frac{1}{2}[\partial_\tau-\partial_{\tau'}+2\mu]\va{\rvek{J}}_1(\tau,\tau')\edot
\end{equation}
Using this formula only the matrix elements of the current operator are needed, which do not include the impurity potential.

\section{Current operators under external fields}
\label{app:currex}
Following Luttinger\cite{Luttinger1964} the Hamiltonian of the $\mathcal{H}_0$ system under external fields can be written as:
\begin{equation}
\mathcal{H}=\int \D^3x h_0(\vek{x})[1+\psi(\vek{x})]+\varrho_e(\vek{x})\phi(\vek{x})\ecom
\end{equation}
The kernel function in Eq. (\ref{eq:kernel}) of this Hamiltonian is:
\begin{align}
\nonumber
H_{ab}(\vek{x},\vek{x}')=&H_{ab}^{(0)}(\vek{x},\vek{x}')\left[1+\frac{1}{2}(\psi(\vek{x})+\psi(\vek{x}'))\right]+\\&+\delta(\vek{x}-\vek{x}')\varrho_e(\vek{x})\phi(\vek{x})\edot
\end{align}
Using Eqs. (\ref{eq:currker}) and (\ref{eq:qcurrker}) the single-particle current and energy current operators in Eq. (\ref{eq:currdiff}) can be expressed as:
\begin{subequations}
\begin{align}
\label{eq:Jtot0}
\va{J}^{\mathrm{tot}}_1=&~\va{J}_1+\frac{1}{2}\left[\va{J}_1\psi+\psi\va{J}_1\right]\ecom\\
\nonumber
\va{J}^{\mathrm{tot}}_2=&~\va{J}_{2}-\frac{e}{2}\left[\va{J}_1\phi+\phi\va{J}_1\right]+\\
\label{eq:Jtot1}
&+\frac{1}{2}\left[\va{J}_{2}\psi+\psi\va{J}_{2}+\va{J}_1\psi\vek{H}_0+\vek{H}_0\psi\va{J}_1\right]\ecom
\end{align}
\end{subequations}
where only the first order terms in the external fields are kept and
\begin{align}
\va{J}_1&=\grad_{\vek{p}}\vek{H}_0\ecom &
\va{J}_{2}&=\frac{1}{2}[\va{J}_{1}\vek{H}_0+\vek{H}_0\va{J}_1]\edot
\end{align}
This is equivalent to the currents obtained in Refs. \onlinecite{Smrcka1977} and \onlinecite{Qin2011}.

\section{Transport coefficients of a general multi-band Hamiltonian}
\label{app:trans}
In this appendix, we prove Eqs. (\ref{eq:appexplained}), (\ref{eq:L11full}) and (\ref{eq:appexplained2}) for a general multi-band Hamiltonian without interactions. The following results are similar to that of Smrčka and Středa\cite{Smrcka1977} who studied a single-band Hamiltonian, but expressed in the eigenstate basis.
Phenomenologically the current density and energy current density can be expressed using the transport coefficients ($\mat{L}_{ij}$) as\cite{Smrcka1977,Luttinger1964,Mahan2000}:
\begin{subequations}
  \begin{align} 
    \vek{j}_1 &= -e^2\vek{L}_{11}\grad\phi+e\vek{L}_{12}\grad\psi\ecom\\
    \vek{j}_2 &= \phantom{-}e\phantom{^2}\vek{L}_{21}\grad\phi  - \phantom{e}\vek{L}_{22}\grad\psi\ecom
  \end{align}
\end{subequations}
where:
\begin{align}
\vek{j}_1&=-e\frac{\expval{\va{\rvek{J}}_1^{\mathrm{tot}}}}{V}\ecom&
\vek{j}_2&=\frac{\expval{\va{\rvek{J}}_2^{\mathrm{tot}}}}{V}\edot
\end{align}
For uniform electric field the electric potential is:
\begin{align}
\phi &= -\vek{x}\vek{E}\edot
\end{align}
According to the arguments in Ref. \onlinecite{Luttinger1964,Smrcka1977} the gradient of the gravitational potential is equivalent to the the temperature gradient:
\begin{equation}
\grad\psi \equiv -T\grad\left(\frac{1}{T}\right)\edot
\end{equation}
If this is also uniform then:
\begin{equation}
\psi \equiv -T\vek{x}\grad\left(\frac{1}{T}\right)\edot
\end{equation}

Using Eqs. (\ref{eq:Jtot0}), (\ref{eq:Jtot1}) and (\ref{eq:currdiff}) the thermal average of the many-body current operators can be expressed as:
\begin{subequations}
\begin{align}
\expval{\va{\rvek{J}}_1^{\mathrm{tot}}}&=\expval{\va{\rvek{J}}_1}+\expval{\va{\rvek{J}}_1^\psi}_0\ecom\\
\expval{\va{\rvek{J}}_2^{\mathrm{tot}}}&=\expval{\va{\rvek{J}}_2}+\expval{\va{\rvek{J}}_1^\phi}_0+\expval{\va{\rvek{J}}_2^\psi}_0\edot
\end{align}
\end{subequations}
The $\expval{}_0$ is the thermal average using only the $\vek{H}_0$ Hamiltonian. In this formulas we only consider the potentials up to linear order. We can divide the contributions to the transport coefficients coming from the field independent and field dependent currents as:
\begin{align}
\label{eq:LKM}
\vek{L}_{ij}=\vek{K}_{ij}+\vek{M}_{ij}\edot
\end{align}

The $\vek{K}_{ij}$ components coming from the $\expval{\va{\rvek{J}}_i}$ terms can be calculated using the Kubo response theory\cite{Mahan2000} as:
\begin{subequations}
\begin{align}
\mat{K}_{ij} &= \lim\limits_{\omega\to0}\frac{i}{\omega}\lim\limits_{\delta\to0^+}\vb*{\Pi}_{ij}(i\omega_\lambda=\omega+i\delta)\ecom\\
\Pi^{(ij)}_{\alpha\beta}(i\omega_\lambda) &= -\frac{1}{V}\int\limits_0^\beta \D\tau\ead{i\omega_\lambda\tau}\expval{\mathcal{J}^{(i)}_\alpha(\tau)\mathcal{J}^{(j)}_\beta(0)}_0\edot
\end{align}
\end{subequations}
Using the many-body current operators with the formalism described in Eq. (\ref{eq:currmahan}) the current-current correlation can be calculated as:
\begin{widetext}
\begin{subequations}
\begin{align}
\label{eq:correxpv}
  \Pi^{(ij)}_{\alpha\beta}(i\omega_\lambda) = -\frac{1}{V}\int\limits_0^\beta \D\tau\ead{i\omega_\lambda\tau}\lim\limits_{\substack{\tau'\to\tau^-\\\tau'''\to\tau''^-\\\tau''\to0^-}}\Delta_i(\partial_\tau,\partial_{\tau'})\Delta_j(\partial_{\tau''},\partial_{\tau'''})\expval{\mathcal{J}_\alpha(\tau,\tau')\mathcal{J}_\beta(\tau'',\tau''')}_o\ecom
\end{align}
\begin{align}
\Delta_1(\partial_\tau,\partial_{\tau'})&\defeq1 \ecom&
\Delta_2(\partial_\tau,\partial_{\tau'})&\defeq\frac{1}{2}(\partial_\tau-\partial_{\tau'}+2\mu)\edot
\end{align}
\end{subequations}
Using the (\ref{eq:manybodyJ}) form of the many-body current operator, performing the thermal average over the field operators and transforming to the Matsubara frequency space we get:
\begin{align}
\Pi^{(ij)}_{\alpha\beta}(i\omega_\lambda) &= \frac{1}{V}  \frac{1}{\beta}\sum\limits_n \int\D^3x\int\D^3x'\Delta_i(i\omega_n,-i\omega_n-i\omega_\lambda)\Delta_j(i\omega_n+i\omega_\lambda,-i\omega_n)\Tr{ \vek{J}_\alpha \mat{G}(\vek{x},\vek{x}',i\omega_n+i\omega_\lambda) \mat{J}_\beta \mat{G}(\vek{x}',\vek{x},i\omega_n)}\ecom
\end{align}
where $\vek{G}$ is the Green's function of the $\vek{H}_0$ Hamiltonian. In the eigenstate basis ($\vek{H}_0\ket{a}=E_a\ket{a}$) this can be expressed as:
\begin{subequations}
\begin{align}
\Pi^{(ij)}_{\alpha\beta}(\omega)= -\frac{1}{V}  \sum\limits_{a,b} J^{(\alpha)}_{ab} J^{(\beta)}_{ba} C^{(ij)}_{ba}(\omega)\ecom
\end{align}
\begin{align}
  C^{(ij)}_{ba}(\omega)\defeq 2\int\limits_{-\infty}^\infty \frac{\D\varepsilon}{2\pi}\left(\varepsilon+\frac{1}{2}\omega\right)^{i+j-2}\bigg[f(\varepsilon-\mu)~G_b^R(\varepsilon+\omega)\Im G^R_a(\varepsilon)+f(\varepsilon-\mu+\omega)~\Im G^R_b(\varepsilon+\omega)G^A_a(\varepsilon)\bigg]\ecom
\end{align}
\end{subequations}
where the Matsubara summation was substituted to an integral\cite{Bruus2004,Abrikosov1988}. The retarded and advanced Green's functions are defined as $G_a^{R/A}\defeq(\varepsilon\pm i\delta-E_a)^{-1}$. After performing the $\omega\to0$ limit we get:
\begin{subequations}
\begin{equation}
  \Re K^{(ij)}_{\alpha\beta} = \frac{1}{V}\sum\limits_{a,b} \Im{J^{(\alpha)}_{ab} J^{(\beta)}_{ba}}\Re C^{(ij)}_{ba}+\Re{J^{(\alpha)}_{ab} J^{(\beta)}_{ba}}\Im C^{(ij)}_{ba}\ecom
\end{equation}
\begin{equation}
  C^{(ij)}_{ba} = 2\int\limits_{-\infty}^\infty \frac{\D\varepsilon}{2\pi} \varepsilon^{i+j-2}\bigg(f(\varepsilon-\mu)\left[\partial_\varepsilon G_b^R(\varepsilon)\Im G^R_a(\varepsilon)+\partial_\varepsilon\Im G^R_b(\varepsilon)G^A_a(\varepsilon)\right]+\partial_\varepsilon f(\varepsilon-\mu)~\Im G^R_b(\varepsilon)G^A_a(\varepsilon)\bigg)\edot
\end{equation}
\end{subequations}
Since the one-particle current operator is hermitian the following relations hold:
\begin{align}
\Im{J^{(\alpha)}_{ab}J^{(\beta)}_{ba}}&=-\Im{J^{(\alpha)}_{ba}J^{(\beta)}_{ab}}\ecom &
\Re{J^{(\alpha)}_{ab}J^{(\beta)}_{ba}}&=\Re{J^{(\alpha)}_{ba}J^{(\beta)}_{ab}}\edot
\end{align}
Using these and partial integrations:
\begin{subequations}
\begin{align}
  \Im C^{(ij)}_{ba} &= -2\int\limits_{-\infty}^\infty \frac{\D\varepsilon}{2\pi} \varepsilon^{i+j-2}\partial_\varepsilon f(\varepsilon-\mu)~\Im G^R_b(\varepsilon) \Im G^R_a(\varepsilon)\ecom\\
 \Re C^{(ij)}_{ba} &= 2\int\limits_{-\infty}^\infty \frac{\D\varepsilon}{2\pi} f(\varepsilon-\mu)\bigg[2\varepsilon^{i+j-2}\partial_\varepsilon \Re G_b^R(\varepsilon)\Im G^R_a(\varepsilon)+(i+j-2)\varepsilon^{i+j-3}\Re G_b^R(\varepsilon)\Im G^R_a(\varepsilon)\bigg]\edot
\end{align}
\end{subequations}
\end{widetext}

\pagebreak ~\\
\pagebreak ~\\

We move on with expressing the $\vek{M}_{ij}$ components in Eq. (\ref{eq:LKM}). Using the eigenstate representation we can write them as:
\begin{subequations}
\begin{align}
\vek{M}_{11} &= \vek{0}\ecom\\
\vek{M}_{12} &= -2\sum\limits_a\int\limits_{-\infty}^\infty \frac{\D\varepsilon}{2\pi} f(\varepsilon-\mu)\Im{G^R_a(\varepsilon)}\mat{M}_{aa}\ecom\\
\vek{M}_{21} &= \vek{M}_{12}\ecom\\
\vek{M}_{22}&=-4\sum\limits_a\int\limits_{-\infty}^\infty \frac{\D\varepsilon}{2\pi} \varepsilon f(\varepsilon-\mu)\Im{G^R_a(\varepsilon)}\mat{M}_{aa}
\ecom
\end{align}
\end{subequations}
where:
\begin{equation}
M^{\alpha\beta}_{aa}=\frac{1}{2}\bra{a}\left[\vb{J}_\alpha x_\beta+x_\beta \vb{J}_\alpha\right]\ket{a}\edot
\end{equation}
Using $\vek{J}_\alpha=i[\vek{H}_0,x_\alpha]$ this can be transformed to:
\begin{equation}
M^{\alpha\beta}_{aa}=\frac{1}{2}\bra{a}\left[\vb{J}_\alpha x_\beta-\vb{J}_\beta x_\alpha \right]\ket{a}\edot
\end{equation}
From the above formula we can see that $M^{ij}_{\alpha\alpha}=0$.

Since every formula is proportional to $f(\varepsilon-\mu)$ or $\partial_\varepsilon f(\varepsilon-\mu)$ it is always possible to express the finite temperature quantities with the zero temperature quantities as:
\begin{equation}
\mat{L}_{ij}(T,\mu)=-\int\dd \varepsilon \dv{f(\varepsilon-\mu)}{\varepsilon}\mat{L}_{ij}(0,\varepsilon)\ecom
\end{equation}
With this the diagonal components are:
\begin{subequations}
\begin{align}
L^{11}_{\alpha\alpha}(0,\varepsilon)&=\frac{1}{\pi V}\sum\limits_{a,b} \left\lvert J^{(\alpha)}_{ab}\right\rvert^2 \Im G^R_b(\varepsilon) \Im G^R_a(\varepsilon)\ecom\\
L^{12}_{\alpha\alpha}(0,\varepsilon)&=\varepsilon L^{11}_{\alpha\alpha}(0,\epsilon)\ecom\\
L^{22}_{\alpha\alpha}(0,\varepsilon)&=\varepsilon^2 L^{11}_{\alpha\alpha}(0,\epsilon)\edot
\end{align}
\end{subequations}
For the off-diagonal components the calculation is more complex. We need to address both $\vek{K}$ and $\vek{M}$ contributions. It can be show to a similar fashion as Ref. \onlinecite{Smrcka1977} that the components coming from $\vek{M}$ compensate for terms coming from $\vek{K}$ in a way that the off-diagonal components can be expressed similarly to the diagonal components as:
\begin{subequations}
\label{eq:generalL}
\begin{align}
\nonumber
L^{11}_{\alpha\beta}(0,\varepsilon)&=\frac{1}{\pi V}\sum\limits_{a,b} \bigg[ \Re{J^{(\alpha)}_{ab}J^{(\beta)}_{ba}}\Im G^R_b(\varepsilon) \Im G^R_a(\varepsilon) +\\ &+ \Im{J^{(\alpha)}_{ab}J^{(\beta)}_{ba}}\int\limits_{-\infty}^\varepsilon \D\xi 2\partial_\xi \Re G_b^R(\xi)\Im G^R_a(\xi)\bigg]\ecom\\
L^{12}_{\alpha\beta}(0,\varepsilon)&=\varepsilon L^{11}_{\alpha\beta}(0,\epsilon)\ecom\\
L^{22}_{\alpha\beta}(0,\varepsilon)&=\varepsilon^2 L^{11}_{\alpha\beta}(0,\epsilon)\edot
\end{align}
\end{subequations}

\section{Calculation of vertex correction in magnetic field}

\label{app:vertex}
In the previous section we derived the formula for the transport coefficients (Eq. (\ref{eq:generalL})). This formula assumes that the eigenstate representation and the Green's function of the whole Hamiltonian is known. In the case of impurities usually the clean system is solvable and we treat the impurities as perturbation. We assume that the Hamiltonian has the form:
\begin{equation}
\vek{H}=\vek{H}_D+\sum\limits_i u(\vek{x}-\vek{x}_i)\edot
\end{equation}
Now the eigenstate basis will be defined using the eigenstates of $\vek{H}_D$ (for simplicity we denote them with the same index as before $\vek{H}_D\ket{a}=E_a\ket{a}$).

Here we only consider the first order approximation in the impurity density. After resummations\cite{Bruus2004} diagrammatically this can be represented as in Fig. \ref{fig:feyn_pi}. The double lines in the diagrams represent the impurity Green's function (which is assumed to be diagonal in the $\ket{a}$ basis). We will only discuss the diagonal components (the off-diagonal components generally give finite results at the zeroth order approximation). 

\fig{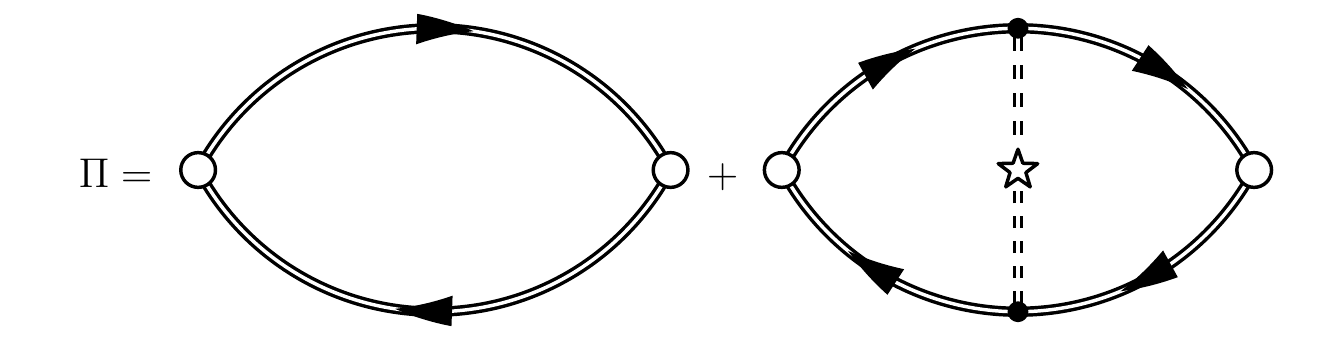}{width=.45\textwidth}{Feynman diagrams of the first order approximation of the correlation function. The double lines are the impurity Green's function, the double dashed lines are the effective impurity potentials and the star is the impurity density.}
The first diagram gives exactly the same contribution as in Eq. (\ref{eq:generalL}), but now the expression is in the eigenstate basis of only $\vek{H}_D$ and the Green's functions are the impurity Green's functions:
\begin{align}
L^{11(0)}_{\alpha\alpha}(\mu)&=\frac{1}{\pi V}\sum\limits_{a,b} \left\lvert J^{(\alpha)}_{ab}\right\rvert^2 \Im G^R_b(\mu) \Im G^R_a(\mu)\edot
\end{align}

The second diagram in the momentum representation can be expressed as:
\begin{widetext}
\begin{equation}
\Pi^{(1)}_{\alpha\beta}(i\omega_{\lambda})=\frac{1}{V^6}\sum\limits_{\substack{\vek{k},\vek{k'}\\ \vek{k''},\vek{k'''}\\ \vek{q}}} \frac{1}{\beta}\sum\limits_nn_i|u_{\vek{q}}|^2 \Tr{\vek{J}^{(\alpha)}_{\vek{kk}}\vek{G}_{\vek{kk'''}}(i\omega_n)\vek{G}_{\vek{k'''-qk''}}(i\omega_n)\vek{J}^{(\beta)}_{\vek{k''k''}}\vek{G}_{\vek{k''k'-q}}(i\omega_n+i\omega_\lambda)\vek{G}_{\vek{k'k}}(i\omega_n+i\omega_\lambda)}\edot
\end{equation}
In the eigenstate basis this becomes:
\begin{subequations}
\begin{align}
\Pi^{(1)}_{\alpha\beta}(i\omega_\lambda)&=\frac{1}{V}\sum\limits_{a,b,c,d} V_{ab}^{dc}{J}^{(\alpha)}_{ad}{J}^{(\beta)}_{cb}C_{ab}^{dc}(i\omega_\lambda)\ecom\\
C_{ab}^{dc}(i\omega_\lambda)&\defeq\frac{1}{\beta}\sum\limits_n G_a(i\omega_n+i\omega_\lambda)G_{b}(i\omega_n+i\omega_\lambda)G_{c}(i\omega_n)G_{d}(i\omega_n)\ecom\\
V_{ab}^{dc}&\defeq\frac{1}{V^3}\sum\limits_{\substack{\vek{k},\vek{k'}\\ \vek{q}}}n_i|u_{\vek{q}}|^2 \gvek{\phi}_d^\dag(\vek{k'})\gvek{\phi}_c(\vek{k'-q})\gvek{\phi}_b^\dag(\vek{k-q})\gvek{\phi}_a(\vek{k})\edot
\end{align}
\end{subequations}
After analytic continuation and expressing the Matsubara sum with an integral:
\begin{align}
C_{ab}^{dc}(\omega)&=-\int\limits_{-\infty}^{\infty}\frac{\D \varepsilon}{2\pi i}f(\varepsilon)\left[g_{abcd}^{RR}(\varepsilon+\omega,\varepsilon)-g_{abcd}^{RA}(\varepsilon+\omega,\varepsilon)+g_{abcd}^{RA}(\varepsilon,\varepsilon-\omega)-g_{abcd}^{AA}(\varepsilon,\varepsilon-\omega)  \right]\ecom\\
g^{XX'}_{abcd}(\varepsilon,\varepsilon')&\defeq G^X_a(\varepsilon)G^X_{b}(\varepsilon)G^{X'}_{c}(\varepsilon')G^{X'}_{d}(\varepsilon')\edot
\end{align}
We will assume that everything except $C_{abcd}(\omega)$ is real (it is not necessarily true but it is true in our case). With this after performing the DC-limit the vertex correction to the conductivity at zero temperature becomes:
\begin{subequations}
\begin{align}
L_{\alpha\alpha}^{11(1)}(\mu)&=\frac{1}{\pi V}\sum\limits_{a,b,c,d} V_{ab}^{dc}{J}^{(\alpha)}_{ad}{J}^{(\alpha)}_{cb}\Im C_{ab}^{dc}(\mu)\ecom\\
\Im C_{ab}^{dc}(\mu)&=\frac{[ (E_a-\mu)\Gamma_b+(E_b-\mu)\Gamma_a][(E_c-\mu)\Gamma_d +(E_d-\mu)\Gamma_c]}{\Gamma_a\Gamma_b\Gamma_c\Gamma_d}\Im G_a^R(\mu) \Im G_b^R(\mu) \Im G_c^R(\mu) \Im G_d^R(\mu)\ecom
\end{align}
\end{subequations}
where $\Gamma_a\equiv\Gamma_a(\mu)$.

\end{widetext}
\bibliography{bibliography.bib}

\end{document}